\documentclass[PAPER, atlasdraft=false, texlive=2025, UKenglish, cernpreprint, texmf, orcidlogo]{atlasdoc} %
\usepackage{atlaspackage}
\usepackage{atlasbiblatex}

\usepackage{atlasphysics}
\graphicspath{{logos/}{figures/}}

\usepackage{ANA-EXOT-2022-13-PAPER-defs}

\AtlasTitle{Search for $t\bar{t}H/A \rightarrow t\bar{t}t\bar{t}$ production in proton--proton collisions at $\sqrt{s}=13$~\TeV\ with the ATLAS detector}

\AtlasAbstract{%
A search is presented for a heavy scalar ($H$) or pseudo-scalar ($A$) predicted by the
two-Higgs-doublet models, where the $H/A$ is produced in association with
a top-quark pair ($t\bar{t}H/A$), and with the $H/A$ decaying into a $t\bar{t}$ pair.
The full LHC Run~2 proton--proton collision data collected by the ATLAS experiment is used,
corresponding to an integrated luminosity of 139~fb$^{-1}$.
Events are selected requiring exactly one or two opposite-charge electrons or muons.
Data-driven corrections are applied to improve the modelling of the $t\bar{t}$+jets background
in the regime with high jet and $b$-jet multiplicities.
These include a novel multi-dimensional kinematic reweighting based on a neural network trained
using data and simulations.  An $H/A$-mass parameterised graph neural network is trained to
optimise the signal-to-background discrimination.
In combination with the previous search performed by the ATLAS Collaboration in the multilepton final state,
the observed upper limits on the $t\bar{t}H/A \rightarrow t\bar{t}t\bar{t}$ production cross-section
at 95\% confidence level range between 14~fb and 5.0~fb for an $H/A$ with mass between
400~\GeV\ and 1000~\GeV, respectively.
Assuming that both the $H$ and $A$ contribute to the $t\bar{t}t\bar{t}$ cross-section, $\tan\beta$ values below 1.7 or
0.7 are excluded for a mass of 400~\GeV\ or 1000~\GeV, respectively.
The results are also used to constrain a model predicting the pair production of
a colour-octet scalar, with the scalar decaying into a $t\bar{t}$ pair.
}

\AtlasRefCode{EXOT-2022-13}
\PreprintIdNumber{CERN-EP-2024-197}

\AtlasDate{\today}

\AtlasJournalRef{\EPJC 85 (2025) 573}
\AtlasDOI{10.1140/epjc/s10052-025-14041-z}

\hypersetup{pdftitle={ATLAS document},pdfauthor={The ATLAS Collaboration}}

\begin{document}

\maketitle

\tableofcontents

%
%
%
%

%
\section{Introduction}
\label{sec:intro}

The discovery of the Higgs boson by the ATLAS and CMS collaborations at the Large Hadron Collider (LHC)
completes the list of elementary particles in the Standard Model (SM)~\cite{HIGG-2012-27,CMS-HIG-12-028}.
Shortcomings of the SM remain, such as quadratic divergences arising from radiative corrections to the Higgs
boson mass~\cite{Susskind:1978ms} and the nature of non-baryonic dark matter~\cite{Bertone_2018}. As a result, many beyond the
Standard Model (BSM) theories have been proposed with extended Higgs sectors, such as
two-Higgs-doublets models (2HDM)~\cite{Branco:2011iw},
hMSSM~\cite{Djouadi:2015jea,Maiani:2013nga,Djouadi:2013uqa}, and models with
Higgs boson triplets~\cite{Lee:1973iz,Cheng:1980qt,Schechter:1980gr,Lazarides:1980nt,Mohapatra:1980yp,Magg:1980ut}.
A search is presented for additional heavy Higgs bosons using a four-top-quark
(\tttt) final state.
In Type-II 2HDM models, the Higgs boson sector consists of two complex doublets of
Higgs boson fields and generates five scalar states including CP-even neutral Higgs
bosons $h$ and $H$, a CP-odd neutral Higgs boson, $A$, and two charged Higgs boson states $H^{\pm}$.
There are five free parameters including three Higgs boson masses, the mixing angle
of the two CP-even Higgs boson states ($\alpha$) and the ratio of the vacuum
expectation values of the Higgs boson doublets (\tanb).
In what is called the  `alignment limit',
where $\sin(\beta-\alpha) = 1$, the $h$ couplings in 2HDM
matches those of the SM Higgs boson.

Direct searches for heavy neutral bosons were carried out by the ATLAS and
CMS collaborations~\cite{HDBS-2018-46,CMS-HIG-17-020,HDBS-2018-58,CMS-HIG-20-014,EXOT-2020-25,CMS-HIG-17-027,EXOT-2019-26,CMS-TOP-18-003},
and also indirect searches using
precision measurements of the production cross-sections and decay rates of the SM Higgs
boson~\cite{HIGG-2015-03}.
These searches and measurements effectively restrict the available parameter space to the
alignment limit~\cite{HIGG-2022-17}.

In the 2HDM model, for $H/A$ masses $> 500$~\GeV\ and modest values of \tanb,
the predominant decay mode is into a \ttbar\ pair.
Searches for \HAtt\ produced via a gluon--gluon initiated loop are difficult because
of destructive interference
with the SM \ttbar\ background that distorts the Breit--Wigner peak in the \ttbar\
mass spectrum~\cite{Hajer:2015gka,Gaemers:1984sj}. Both the ATLAS experiment (20.3~\ifb\ at $\sqrt{s} = 8$~\TeV)
and the CMS experiment (36~\ifb\ at $\sqrt{s} = 13$~\TeV) searched for  $gg\rightarrow A/H \rightarrow \ttbar$
taking the interference term into account~\cite{EXOT-2016-04, CMS-HIG-17-027}.
The ATLAS experiment recently
published results in the same channel using the full Run 2 data set at $\sqrt{s} = 13$~\TeV~\cite{EXOT-2020-25}.
Alternatively, $H/A$ production in association with a \ttbar\ pair, followed by \HAtt,
is much less susceptible to interference effects with the SM \tttt\ production.
An illustrative Feynman diagram for \ttHAtt\ production is shown in Figure~\ref{fig:Feynman}.
Searches for $H/A$ production in the \tttt\ final state are therefore promising,
especially given that the SM \tttt\ production has now been observed
by both the ATLAS~\cite{TOPQ-2021-08} and CMS experiments~\cite{CMS-TOP-22-013}.
The measured cross-sections
using final states with two leptons with the same electric charge or at least three leptons (\SSML)
are $22.5^{+4.7}_{-4.3} (\mathrm{stat.}) ^{+4.6}_{-3.4} (\mathrm{syst.})$~fb by the ATLAS experiment
and $17.7 ^{+3.7}_{-3.5} (\mathrm{stat.}) ^{+2.3}_{-1.9} (\mathrm{syst.})$~fb by the CMS experiment.
Both of the measurements are consistent with the latest SM prediction of
$13.4^{+1.0}_{-1.8}$~fb,
computed at next-to-leading-order (NLO) with full electroweak (EW) corrections
including threshold resummation at next-to-leading-logarithm (NLL$^\prime$) accuracy~\cite{vanBeekveld:2022hty}.
With the present uncertainties, both the measurements leave room for additional contributions to
SM \tttt\ production.
Both the ATLAS and CMS experiments have searched for \ttHAtt\ production in the \SSML\ final state at
$\sqrt{s} = 13$~\TeV~\cite{EXOT-2016-16,EXOT-2019-26,CMS-SUS-16-035,CMS-TOP-18-003}.

\begin{figure}[tbp!]
\centering
\includegraphics[width=0.5\linewidth, ]{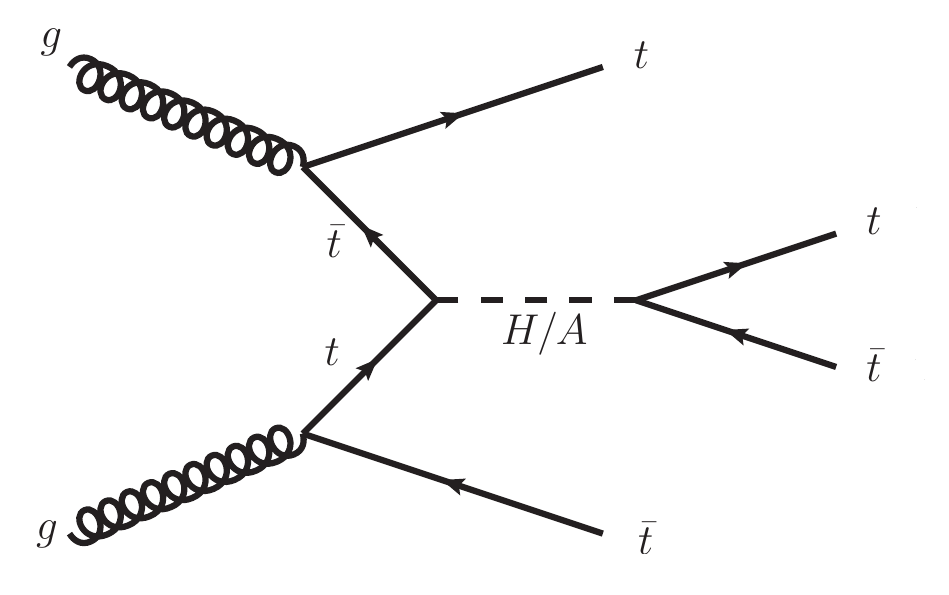}
\caption{Illustrative Feynman diagram showing the production of a heavy scalar or pseudo-scalar Higgs boson, $H/A$,
produced in association with a pair of top quarks and the Higgs boson decaying into a pair of top quarks.}
\label{fig:Feynman}
\end{figure}

A search is presented for
\ttHAtt\ in the $H/A$ mass ($m_H$) range of $400-1000$~\GeV.
The search uses events with exactly one lepton (electron or muon) or
two leptons with opposite electric charge (\SLOS), and multiple heavy-flavour jets
selected from the 139~\ifb\ of proton--proton ($pp$) collision data at $\sqrt{s} = 13$~\TeV\
collected by the ATLAS experiment during 2015--2018.
The analysis adopts similar baseline strategies to those used in
the \tttt\ cross-section measurement in the \SLOS\ channel~\cite{TOPQ-2020-10},
with improved physics object reconstruction and identification,
advanced background modelling of \ttjets\ with multivariate data-driven corrections,
and a new $m_{H/A}$-parameterised multivariate discriminant.
The cross-section of SM \tttt\ production is taken to be the latest NLO+NLL$^\prime$ prediction from Ref.~\cite{vanBeekveld:2022hty} in the search.
Due to the smallness of the SM \tttt\ production cross-section
and the little discrimination between SM \tttt\ and \ttHAtt\ production with $m_{H/A}$ close to 400~\GeV,
a simultaneous measurement of the SM \tttt\ cross-section is infeasible.
Limits are produced on the \ttHA\ production cross-section times
branching ratio of \HAtt\ as a function of $m_{H/A}$.
A combination with a previous search using the \SSML\ final state
and the same data sample~\cite{EXOT-2019-26} is performed.
The upper limits on the cross-sections are interpreted in the context of a Type-II 2HDM
in the alignment limit.
The analysis also sets limits on a model predicting the pair production of
a colour-octet scalar, usually known as a sgluon,
where the sgluons decay into \ttbar\ pairs~\cite{Darm__2021}.


%
\section{ATLAS detector}
\label{sec:atlas}
\newcommand{\AtlasCoordFootnote}{%
ATLAS uses a right-handed coordinate system with its origin at the nominal interaction point (IP)
in the centre of the detector and the \(z\)-axis along the beam pipe.
The \(x\)-axis points from the IP to the centre of the LHC ring,
and the \(y\)-axis points upwards.
Polar coordinates \((r,\phi)\) are used in the transverse plane,
\(\phi\) being the azimuthal angle around the \(z\)-axis.
The pseudorapidity is defined in terms of the polar angle \(\theta\) as \(\eta = -\ln \tan(\theta/2)\) and is equal to the rapidity
$ y = \frac{1}{2} \ln \left( \frac{E + p_z c}{E - p_z c} \right) $ in the relativistic limit.
Angular distance is measured in units of \(\Delta R \equiv \sqrt{(\Delta y)^{2} + (\Delta\phi)^{2}}\).}

The ATLAS detector~\cite{PERF-2007-01} at the LHC covers nearly the entire solid angle around the collision point.\footnote{\AtlasCoordFootnote}
It consists of an inner tracking detector surrounded by a thin superconducting solenoid, electromagnetic and hadronic calorimeters,
and a muon spectrometer incorporating three large superconducting air-core toroidal magnets.

The inner-detector system (ID) is immersed in a \qty{2}{\tesla} axial magnetic field
and provides charged-particle tracking in the range of \(|\eta| < 2.5\).
The high-granularity silicon pixel detector covers the vertex region and typically provides four measurements per track,
the first hit generally being in the insertable B-layer (IBL) installed before Run~2~\cite{ATLAS-TDR-19,PIX-2018-001}.
It is followed by the SemiConductor Tracker (SCT), which usually provides eight measurements per track.
These silicon detectors are complemented by the transition radiation tracker (TRT),
which enables radially extended track reconstruction up to \(|\eta| = 2.0\).
The TRT also provides electron identification information
based on the fraction of hits (typically 30 in total) above a higher energy-deposit threshold corresponding to transition radiation.

The calorimeter system covers the pseudorapidity range of \(|\eta| < 4.9\).
Within the region of \(|\eta|< 3.2\), electromagnetic calorimetry is provided by barrel and
endcap high-granularity lead/liquid-argon (LAr) calorimeters,
with an additional thin LAr presampler covering \(|\eta| < 1.8\)
to correct for energy loss in material upstream of the calorimeters.
Hadronic calorimetry is provided by the steel/scintillator-tile calorimeter,
segmented into three barrel structures within \(|\eta| = 1.7\), and two copper/LAr hadronic endcap calorimeters.
The solid angle coverage is completed with forward copper/LAr and tungsten/LAr calorimeter modules
optimised for electromagnetic and hadronic energy measurements respectively.

The muon spectrometer (MS) comprises separate trigger and
high-precision tracking chambers measuring the deflection of muons in a magnetic field generated by the superconducting air-core toroidal magnets.
The field integral of the toroids ranges between \num{2.0} and \qty{6.0}{\tesla\metre}
across most of the detector.
Three layers of precision chambers, each consisting of layers of monitored drift tubes, cover the region of \(|\eta| < 2.7\),
complemented by cathode-strip chambers in the forward region, where the background is highest.
The muon trigger system covers the range of \(|\eta| < 2.4\) with resistive-plate chambers in the barrel, and thin-gap chambers in the endcap regions.

The luminosity is measured mainly by the LUCID--2~\cite{LUCID2} detector that records Cherenkov light produced in the quartz windows of photomultipliers located close to the beampipe.

Events are selected by the first-level trigger system implemented in custom hardware,
followed by selections made by algorithms implemented in software in the high-level trigger~\cite{TRIG-2016-01}.
The first-level trigger accepts events from the \qty{40}{\MHz} bunch crossings at a rate below \qty{100}{\kHz},
which the high-level trigger further reduces in order to record complete events to disk at about \qty{1}{\kHz}.

A software suite~\cite{SOFT-2022-02} is used in data simulation, in the reconstruction
and analysis of real and simulated data, in detector operations, and in the trigger and data acquisition
systems of the experiment.


%
\section{Object and event preselection}
\label{sec:object_preselection}
The analysis uses $pp$ collision data collected  between 2015 and 2018 by the ATLAS detector  at  $\sqrt{s}=13$~\TeV.
The full data sample corresponds to an integrated luminosity of 139~$\ifb$~\cite{ATLAS-CONF-2019-021} after applying data quality requirements~\cite{DAPR-2018-01}.
Events were collected using single-lepton  triggers, with minimum $\pT$ thresholds  varying from 20 to 26~\GeV  depending on the lepton flavour and the data-taking period~\cite{TRIG-2018-05,TRIG-2018-01}.  Events are required to have a collision vertex matched with at least two ID tracks, each with transverse momentum $\pT>0.5~\GeV$.
The vertex with the largest sum of $\pT^2$ over all matched ID tracks is referred to as the primary vertex~\cite{ATL-PHYS-PUB-2015-026}.

Electron candidates are reconstructed from energy deposits in the electromagnetic calorimeter matched to a track in the ID~\cite{EGAM-2018-01} and are required to be within $|\eta|=2.47$, excluding the calorimeter transition region $1.37<|\eta| < 1.52$.  Muon candidates are reconstructed by combining tracks in the ID with tracks in the MS~\cite{MUON-2018-03} and are required to be within $|\eta|=2.5$.  Both the electron and muon candidates are required to have  $\pT>10~\GeV$.  Electrons must satisfy the `TightLH' identification criterion~\cite{EGAM-2018-01} and are required to be well isolated using criteria based on the sum of the energies of the topological clusters in the calorimeter and \pt\ of ID tracks around the reconstructed electron.  Muons must  meet the `Medium'  identification criterion~\cite{MUON-2018-03} and are required to satisfy isolation requirements based on the \pt\ sum of the ID tracks around the reconstructed muon.  The transverse impact parameter divided by its estimated uncertainty,  $|d_0|/\sigma(d_0)$, is required to be lower than five (three) for electron (muon) candidates. The longitudinal impact parameter must satisfy $|z_0\sin{\theta}|<0.5$~mm for both the lepton flavours.

Jets are reconstructed by combining measurements from both the ID and the calorimeter using a particle-flow algorithm~\cite{PERF-2015-09}.  The anti-$k_t$ algorithm~\cite{Cacciari:2008gp,Fastjet} with a radius parameter of $R=0.4$ is used.  These jets are referred to as `small-$R$ jets' (or jets for simplicity). They are calibrated as described in Ref.~\cite{JETM-2018-05} and  are required to have  $\pT> 25$~\GeV\ and $|\eta|<2.5$.  A  jet-vertex tagger (JVT)~\cite{PERF-2014-03} is applied to jets with $\pT<60$~\GeV\ and $|\eta|<2.4$  to reduce the effect of additional $pp$ collisions in the same or a nearby bunch crossing, collectively referred to as  pile-up. A set of quality criteria are also applied to reject events containing at least one jet arising from non-collision sources or detector noise~\cite{ATLAS-CONF-2015-029}.

Jets containing $b$-hadrons are identified ($b$-tagged) using the DL1r algorithm~\cite{FTAG-2019-07}.
A jet is $b$-tagged if the DL1r score is above a certain threshold, referred to as an operating point (OP).
Four OPs are defined with average  expected efficiencies for $b$-jets of 60\%, 70\%, 77\% and 85\%, as determined in simulated \ttbar\ events.
An integer pseudo-continuous $b$-tagging (pcb) score ranging from five to one is assigned to each jet.
Each score value represents that the jet satisfies one OP but fails to meet the adjacent tighter one.
The largest score of five is assigned to jets satisfying the tightest OP of 60\%.
A score of one is assigned if a jet does not satisfy any of the OPs.

The selected and calibrated small-$R$ jets are used as inputs for jet reclustering~\cite{ATLAS-CONF-2017-062},  using the anti-$k_t$
algorithm with a radius parameter of $R=1.0$. These reclustered jets are referred to as `large-$R$ jets'.
The calibration corrections and uncertainties for the reclustered large-$R$ jets are inherited from the small-$R$ jets.
To suppress the contribution from pile-up, a trimming procedure is applied to the reclustered jets to remove all the matched small-$R$ jets that have \pT below 5\% of the  \pT of the reclustered jet. Large-$R$ jets are required to have $\pT >200$~\GeV\ and $|\eta|<2.0$.

A sequential overlap removal procedure identical to that used in Ref.~\cite{EXOT-2019-26} is applied to ensure that the same calorimeter energy deposit or the same track is not matched with two or more different reconstructed objects.

The missing transverse momentum of the event, whose magnitude is denoted by \met, is defined as the negative vector sum of the $\pT$ of all reconstructed and calibrated objects in the event. This sum includes a term to account for the transverse momenta of ID tracks matched to the  selected primary vertex but not matched with any of the selected objects in the event~\cite{JETM-2020-03}.

Events are preselected by requiring either exactly one lepton (electron or muon) and at least seven jets (\SL\ channel) or exactly two opposite-charge leptons and at least five jets (\OS\ channel).
The highest-$\pT$ (leading) lepton must have $\pT >28$~\GeV.
In the \OS\ channel, the events with two same-flavour leptons must
have a dilepton invariant mass above 15 \GeV and outside the $Z$ boson mass window of 83--99 \GeV.
Events must contain at least two $b$-tagged jets passing the 70\% OP\@.
After the preselection, $\sim$91\% of the background events come from \ttjets\ production,
estimated by using the simulated samples described in Section~\ref{sec:mc} with data-driven corrections discussed in Section~\ref{sec:ttbar_modelling}.


%
\section{Monte Carlo simulations}
\label{sec:mc}

Monte-Carlo (MC) simulated samples were generated for all signal and background processes~\cite{SOFT-2010-01}.
The predictions are based on simulations
produced with \POWHEG~\cite{Nason:2004rx,Frixione:2007nw,Frixione:2007vw,Alioli:2010xd} or \MGNLO~\cite{Alwall:2014hca}
as the matrix element (ME) generator,
interfaced with \PYTHIA[8]~\cite{Sjostrand:2014zea} or \HERWIG[7]~\cite{Bahr:2008pv,Bellm:2015jjp}
for the simulation of parton shower (PS) including fragmentation and hadronisation,
or combined simulations of ME and PS produced with the \SHERPA[2]~\cite{Bothmann:2019yzt} generator
following the \MEPSatNLO\ prescription~\cite{Catani:2001cc,Schumann:2007mg,Hoeche:2009rj,Hoeche:2011fd,Hoeche:2012yf}.
The masses of the top quark and the SM Higgs boson in the relevant samples were set to 172.5 and 125~\GeV, respectively.
The A14 set of tuned parameters (tune)~\cite{ATL-PHYS-PUB-2014-021} was implemented in \PYTHIA[8]
whereas for \HERWIG[7] and \SHERPA\ the generator-specific tuned parameters were used.
Unless otherwise specified, the \NNPDF[2.3lo]~\cite{Ball:2012cx} and \NNPDF[3.0nlo]~\cite{Ball:2014uwa} sets of parton distribution functions (PDF)
were used for samples generated with leading-order (LO) and NLO accuracy in QCD, respectively.
Pile-up events were simulated using \PYTHIA[8.186]~\cite{Sjostrand:2007gs} with the \NNPDF[2.3lo] PDF set
and the A3 tune~\cite{ATL-PHYS-PUB-2016-017},
and overlaid on each hard-scattering event.
\EVTGEN[1.2.0] or 1.6.0~\cite{Lange:2001uf} were used to model heavy-flavour hadron decays in \PYTHIA[8] and \HERWIG[7].
The detector geometry and response were simulated using either \GEANT~\cite{Agostinelli:2002hh}
or a fast simulation (AFII) which parameterises the calorimeter response~\cite{ATL-PHYS-PUB-2010-013}.
All simulated samples were processed through the same reconstruction software as used for data.

The signal samples were simulated
assuming a 2HDM Type-II model in the alignment limit ($\sin(\beta - \alpha) = 1$).
Events of \ttHtt\ were simulated assuming the on-shell production of $H$.
The simulation was performed at LO accuracy in the ME using \MGNLO\ with the \NNPDF[3.1lo]~\cite{Ball:2014uwa} PDF set.
The decay of the top quarks is simulated in the ME generation
to include spin correlations.
The value of $\beta$ was found to yield no change in the event kinematics.
No mixing between $H$ and $A$ was considered.
No dedicated \ttAtt\ samples were simulated
given that the kinematic distributions of interest
differ by less than 1\% between \ttHtt\ and \ttAtt.
Seven signal samples were generated for $m_H$ from 400 to 1000~\GeV\ spaced with 100~\GeV\ steps.
For each assumed $m_H$ value, the $H$ width was set to values consistent with the scenario of $\tanb\sim1$,
ranging from 5~\GeV\ for $m_H=400$~\GeV\ to 30~\GeV\ for $m_H=1000$~\GeV.
Varying the width up to 75\% of $m_H$ yields a change of less than 10\% in only the lepton and
jet \pt\ distributions and negligible changes in the other distributions.
The interference with the SM \tttt\ production was not taken into account.
For the $(\tan\beta, m_{H/A})$ parameter space that this search is sensitive to,
both the effects of the interference and the non-resonant production
are smaller than a few percent, and are therefore neglected.

An additional BSM signal corresponding to the pair production of a colour-octet scalar ($S_8$), a sgluon, was considered,
where both the $S_8$ decay into a \ttbar\ pair.
The production of $S_8S_8$ events was simulated at LO with \MGNLO\
using the simplified model from Ref.~\cite{Darm__2021} with the \NNPDF[3.0nlo] PDF set,
interfaced with \PYTHIA[8].
The decay of $S_8\rightarrow\ttbar$ was handled by \MADSPIN~\cite{Frixione:2007zp,Artoisenet:2012st}.
Simulated samples were generated for $m_{S_8}$ ranging between 0.4~\TeV\ and 2~\TeV, with a spacing
of 100~\GeV\ for $m_{S_8}\in [0.4, 1.5]$~\TeV, and 250~\GeV\ for $m_{S_8}\in [1.75, 2.0]$~\TeV.
The $S_8\ttbar$ coupling parameter $y_{S_8}$ was set to $0.1$, ensuring that $S_8S_8$ production dominates over  $\ttbar S_8$ production.
For $y_{S_8}<0.1$ and $m_{S_8}<2$~\TeV, the interference with the SM background and the $\ttbar S_8$ production
contribute less than 10\% to the total cross-section of \tttt\ production via $S_8$
and were thus not simulated.
The kinematics of $S_8S_8\rightarrow\tttt$ production do not depend on the value of $y_{S_8}$.

The simulated \ttjets\ events are classified into different groups.
This classification is used to construct a detailed model of theoretical systematic uncertainties affecting this process.
The events are classified as \ttbjets, \ttcjets\ and \ttljets\
according to the flavour of the particle jets not originating from top-quark decays.\footnote{
Particle jets are constructed from simulated stable particles (with mean lifetime $\tau > 3 \times 10^{-11}$~s)
using the anti-$k_t$ algorithm with $R=0.4$
and are required to have $\pt>15$~\GeV\ and $|\eta|<2.5$.}.
The flavour and origin of the particle jets are determined
by checking whether any $b$- or $c$-hadron with $\pt > 5$~\GeV\ not originating from top-quark decays
falls within $\dR=0.3$ around the jet axis.
Events with at least one particle jet matched to
such $b$-hadrons are classified as \ttbjets\ events.
Those that are not classified as \ttbjets\ events
but contain at least one particle jet matched to
$c$-hadrons are classified as \ttcjets\ events.
The \ttbjets\ and \ttcjets\ events are collectively referred to as `\ttHF' events (HF for `heavy flavour').
The remaining are classified as \ttljets\ events.
The \ttbjets events are further classified into \ttb, \ttbb, \ttB\ and \ttbbb\ events,
where the $b$ or $\bar{b}$ refers to a particle jet matched to a single $b$-hadron,
and $B$ refers to a particle jet matched to more than one $b$-hadron.
The \ttbbb\ events contain more than two particle jets matched to any number of $b$-hadrons.

The nominal \ttjets\ prediction is based on a sample simulated using \POWHEGBOX[2]
at NLO in QCD using the five-flavour scheme (5FS).
The \hdamp\ parameter,
which regulates the \pt\ of the first additional emission beyond the Born configuration,
is set to 1.5 times the top-quark mass of $m_t=172.5$~\GeV.
The renormalisation and factorisation scales were set to $\frac{1}{2}\Sigma_{i=t,\bar{t}}m_{\mathrm{T},i}$,
where $m_{\mathrm{T},i}$ is the transverse mass of a given parton $i$.
Several samples in addition to the nominal \POWHEGBOX+ \PYTHIA[8] samples were generated to evaluate systematic uncertainties,
including samples generated with alternative ME or PS generators, \MGNLO+ \PYTHIA[8] and \POWHEG+ \HERWIG[7].
Furthermore, a dedicated $\ttbar b\bar{b}$ sample was produced
using \POWHEGBOXRES~\cite{Jezo:2018yaf,ttbbPowheg} and \OPENLOOPS~\cite{Cascioli:2011va,Buccioni:2017yxi,Denner:2016kdg}
with the \NNPDF[3.0nlo\_nf4]~\cite{Ball:2014uwa} PDF set, interfaced with \PYTHIA[8].
This sample was produced using the four-flavour scheme (4FS) with the additional $b$-quarks in the ME generation.
The \hdamp\ parameter was set to $\frac{1}{2}\Sigma_{i=t,\bar{t},b,\bar{b}}m_{\mathrm{T},i}$.
The renormalisation and factorisation scales were set to $\sqrt[4]{\prod_{i=t,\bar{t},b,\bar{b}} m_{\mathrm{T},i}}$
and $\frac{1}{2}\Sigma_{i=t,\bar{t},b,\bar{b},j}m_{\mathrm{T},i}$ (where $j$ stands for extra partons), respectively.
These \ttjets\ samples were previously studied in detail and compared in Ref.~\cite{ATL-PHYS-PUB-2022-026}.

The SM \tttt\ background has an important impact on the search despite its small yield.
The nominal \tttt\ sample was simulated using \MGNLO\ at NLO in QCD with the \NNPDF[3.1nlo] PDF set.
The decay of the top quarks was simulated using \MADSPIN.
The renormalisation and factorisation scales were set to $\frac{1}{4}\Sigma_{i=t,\bar{t},j}m_{\mathrm{T},i}$ following Ref.~\cite{Frederix:2017wme}.
Two alternative samples were generated,
one with \SHERPA[2.2.11],
and another by replacing the \PYTHIA[8]  parton shower  of the nominal sample by \HERWIG[7].
The \SHERPA\ sample was simulated at NLO in QCD\@.
The renormalisation and factorisation scales were set to
half of the scalar sum of the \pt\ of all final state particle.  All \tttt\ samples were normalised to the latest cross-section prediction of
$13.4^{+1.0}_{-1.8}$~fb~\cite{vanBeekveld:2022hty}.

Other backgrounds constitute less than 8\% of the total background.
Most of these events come from \ttH, \ttW, \ttZ, single-top-quark and \Vjets\ productions.
The \ttH\ and $\ttW/Z$ samples were simulated at NLO in QCD using \POWHEGBOX[2] and \MGNLO, respectively.
For the \ttW\ background, a dedicated sample containing only EW corrections was produced at LO in QCD,
and combined with the NLO QCD sample.
Alternative samples for \ttH\ and \ttW\ were simulated using \MGNLO+ \PYTHIA[8] and \SHERPA[2.2.10], respectively.
The production of single-top quarks was modelled using the same generators as the ones used for the inclusive \ttjets\ sample,
including the alternative samples for systematic uncertainties.
The `diagram removal' scheme was used to treat the overlap between \tW\ and \ttbar\ production.
An alternative sample was simulated using the `diagram subtraction' scheme~\cite{Re:2010bp}.
The \Vjets\ samples were simulated using \SHERPA[2.2.1] with the \NNPDF[3.0nnlo] PDF set.
The generation considered ME at NLO in QCD with up to two additional jets and at LO with up to four additional jets.
The remaining events comprise less than 1\% of the total background.
These include events from \ttWW, \tWZ, \tZ, \ttt\ production (simulated using \MGNLO)
and \VV\ production (simulated using \SHERPA[2]).


%
\section{Analysis strategy and event categorisation}
\label{sec:categorisation}

The $t\bar{t}H/A\rightarrow \tttt$ events feature high jet and $b$-jet multiplicities.
The main background affecting the \SLOS final states is \ttjets,
where several additional jets beyond
the \ttbar\ born level are required to give topologies similar to the \tttt signal.
Two known issues affect the modelling of the \ttjets\ background from the simulations described
in Section~\ref{sec:mc}.
The kinematics are mismodelled in the regime of high jet multiplicities
and high \pt\ of the \ttbar\ system~\cite{TOPQ-2018-15} and
the cross-section of \ttHF\ production is underestimated~\cite{ATLAS:2024aht,CMS-TOP-22-009}.
Both of the issues affect the phase space populated by signal events.
Data-driven corrections are adopted to mitigate the mismodelling.
Rescaling factors are derived to correct the fractions and normalizations
of \ttljets, \ttcjets and \ttbjets, referred to as the `flavour rescaling'.
After applying the flavour rescaling,
a neural network (NN) is then trained to correct the kinematic mismodelling.
The details of the data-driven corrections are described in Section~\ref{sec:ttbar_modelling}.
Events passing the preselection are categorised into orthogonal regions.
These include regions used to derive the data-driven corrections,
and the control and signal regions (CR/SR) used in a profile likelihood fit (see Section~\ref{sec:results}) to extract and
set a limit on the signal strength.
In the SRs, a multivariate analysis based on $m_{H/A}$-parameterised graph neural networks (GNN)
is performed to optimise the separation between the signals and the largest background, \ttjets,
detailed in Section~\ref{sec:mva}.

Figure~\ref{fig:categorisation} illustrates the event categorisation.
Events are categorised in bins of jet multiplicity, from seven jets (7j) to at least ten jets ($\geq$10j) in the \SL\ channel
and from five jets (5j) to at least eight jets ($\geq$8j) in the \OS\ channel.
Each jet multiplicity bin is further separated according to the $b$-tagging requirements listed in Table~\ref{tab:categorisation}.
Events are first split according to the number of $b$-tagged jets using the 70\% OP ($N_b^{70\%}$),
from two to at least five (four) in the \SL\ (\OS) channel.
The events with three $b$-tagged jets are further separated into 3bL, 3bH and 3bV regions
given the number of jets passing the 60\% and 85\% OPs ($N_b^{60\%}$ and $N_b^{85\%}$).
The 3bL (3bH) requirement aims to select events with lower (higher) purity of jets originating from $b$-quarks
in events with three jets tagged at the 70\% OP\@.
This is verified in simulated samples using MC generator-level $b$-jets, defined as jets that are matched to $b$-hadrons.
The 3bV requirement is used to define the validation regions (VR).
The resulting six VRs have a signal contamination of $< 2$\%.
The categorisation yields six (five) CRs and six (four) SRs in the \SL\ (\OS) channel.
The events in the 2b regions are used to derive the NN kinematic corrections to the \ttjets\ background.
The events in the 7j (5j) bin in the \SL\ (\OS) channel are used to obtain the \ttjets\ flavour rescaling factors.
These regions used to derive the data-driven corrections all have negligible signal contaminations.
The two sets of corrections are applied to the CRs, VRs and SRs.

\begin{figure}[t!]
\centering
\subfloat[]{\includegraphics[width=0.495\textwidth]{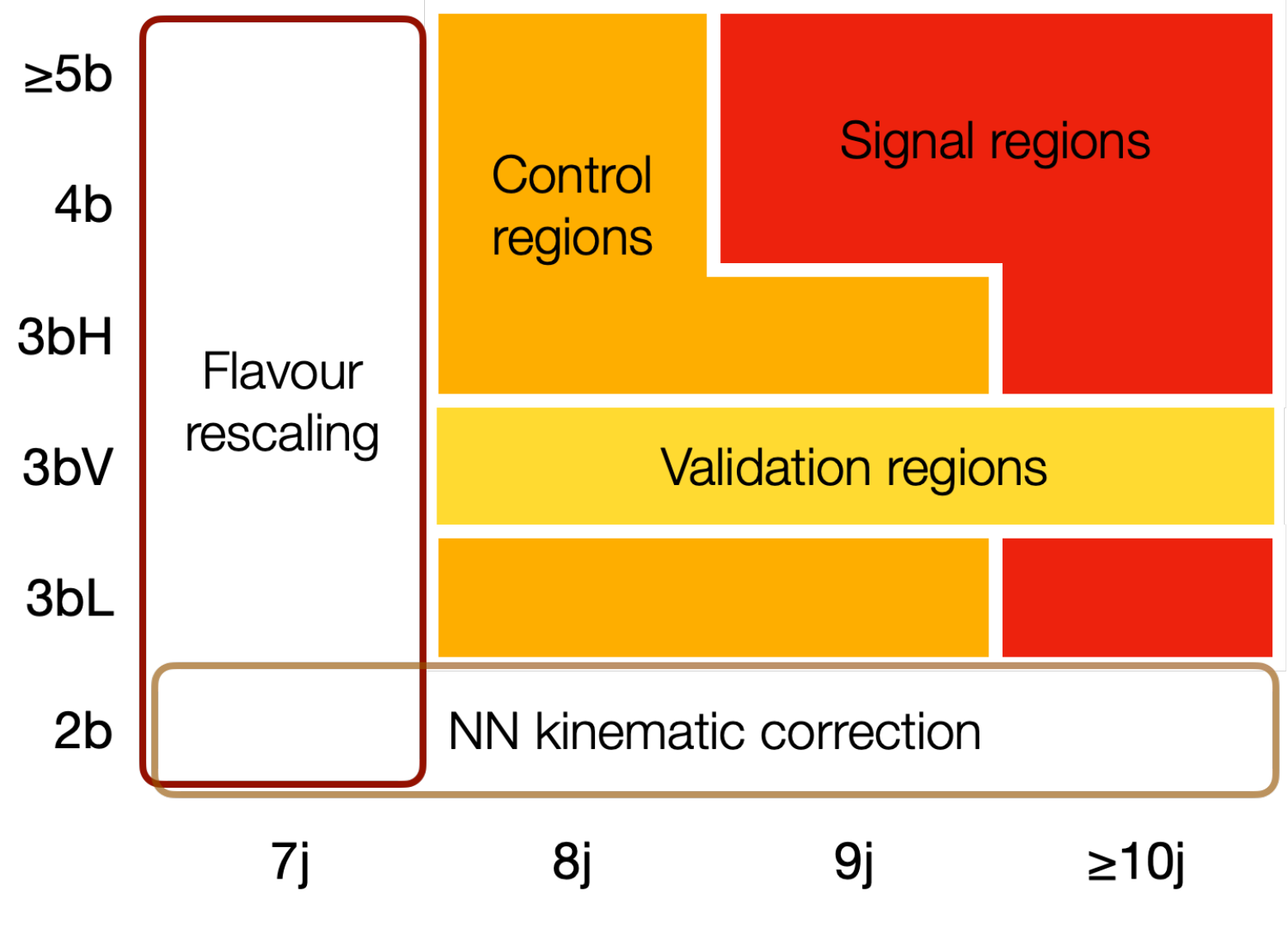}}
\subfloat[]{\includegraphics[width=0.495\textwidth]{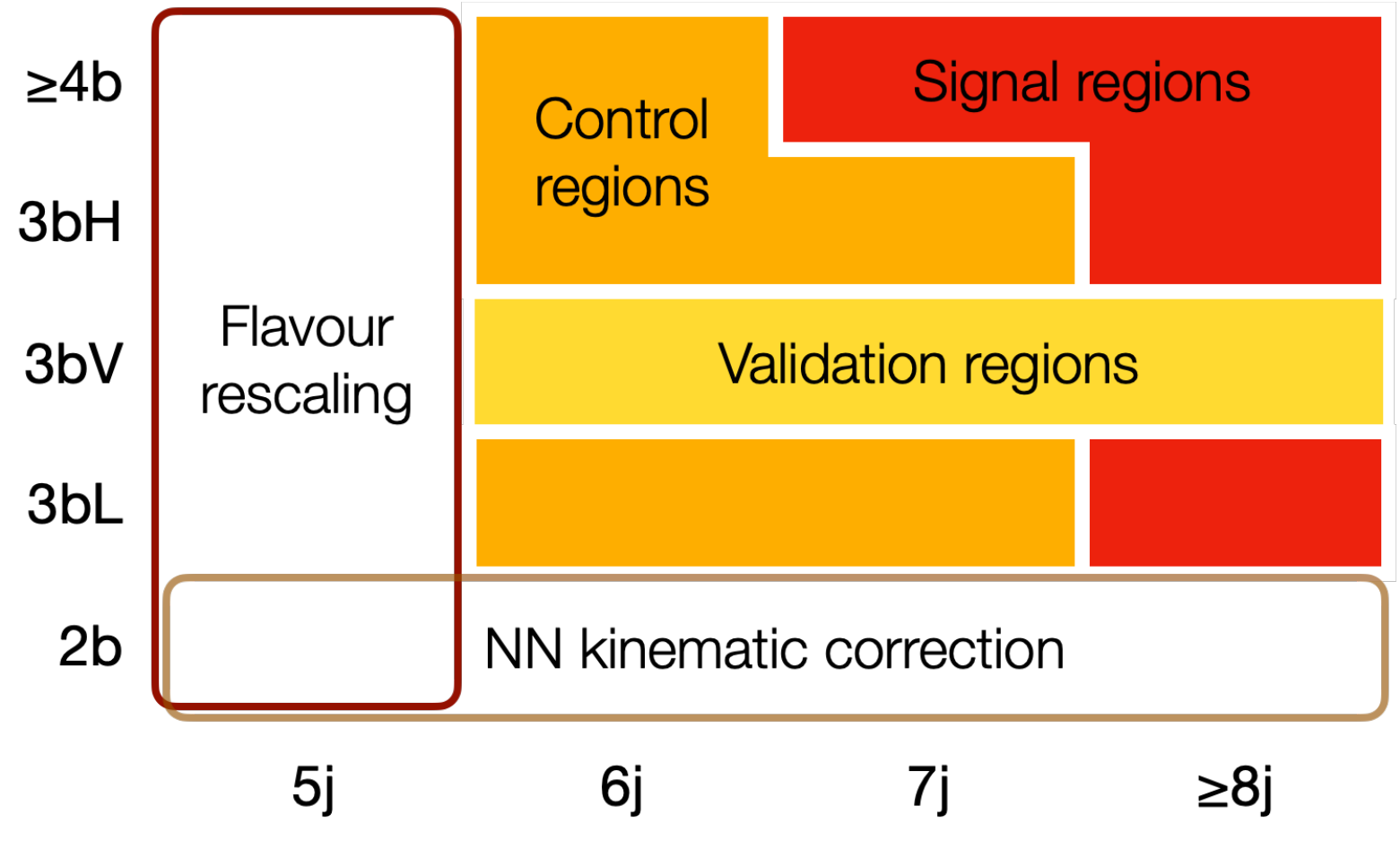}}
\caption{
Schematic view of the event categorisation for the (a) \SL\ and (b) \OS\ channels.
The axes represent the jet multiplicities and $b$-tagging requirements defined in Table~\ref{tab:categorisation}.
Control, signal and validation regions are shaded with different colours.
The regions used to derive the flavour rescaling and NN kinematic corrections
are highlighted using rectangles.
}
\label{fig:categorisation}
\end{figure}

\begin{table}[t!]
\begin{center}
\caption{
Summary of the $b$-tagging requirements used for the event categorisation.
Events in each category must satisfy all requirements listed in the columns.
$N_b^{60\%}$, $N_b^{70\%}$ and $N_b^{85\%}$ are defined as the numbers of $b$-tagged jets
obtained using $b$-tagging operating points with average
expected efficiencies of 60\%, 70\% and 85\%, respectively.
The symbol `--' indicates that no requirement is applied.
\newline
}
\begin{tabular}{m{6em}m{5em}m{5em}m{5em}}
\toprule \midrule
\\[-1em]
Name & $N_b^{60\%}$ &  $N_b^{70\%}$ & $N_b^{85\%}$ \\
\\[-1em]
\midrule
2b             & --       & = 2     & -- \\
3bL            & $\le$ 2 & = 3     & -- \\
3bH            & = 3     & = 3     & > 3 \\
3bV            & = 3     & = 3     & = 3 \\
$\ge$4b (\OS) & --       & $\ge$ 4 & -- \\
4b (\SL)        & --       & = 4     & -- \\
$\ge$5b (\SL)   & --       & $\ge$ 5 & -- \\
\midrule \bottomrule
\end{tabular}
\label{tab:categorisation}
\end{center}
\end{table}

The categorisation using $b$-tagging requirements creates regions enriched with different classes of \ttjets\ events.
In the profile likelihood fit to data, these regions provide constraints on each background class.
The 3bH regions are mainly populated by \ttb, \ttB\ and \ttbb\ events,
whereas the 3bL regions are populated by more \ttljets\ and \ttcjets\ events.
The $\geq$5b regions in the \SL\ channel provide unique constraints on the \ttbbb\ background.
The \OS\ channel has a lower number of events and therefore the $\geq$4b region is not further separated.
The VRs defined using the 3bV requirement are mostly populated by \ttbjets\ events,
which are the most important background events in the SRs.
Figure~\ref{fig:cate_frac} illustrates the background fractions in all the CRs, VRs and SRs.

\begin{figure}[t!]
\centering
\subfloat[]{\includegraphics[width=0.525\textwidth]{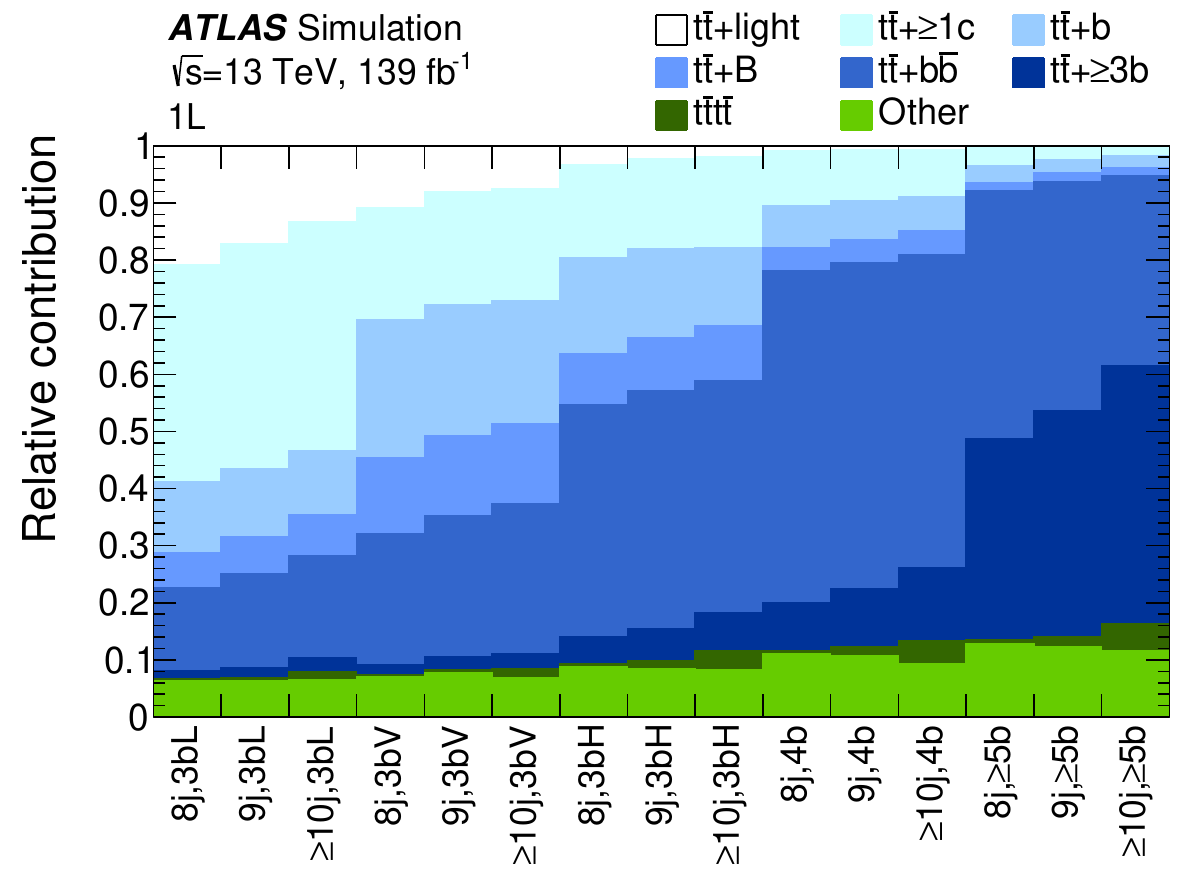}}
\subfloat[]{\includegraphics[width=0.465\textwidth]{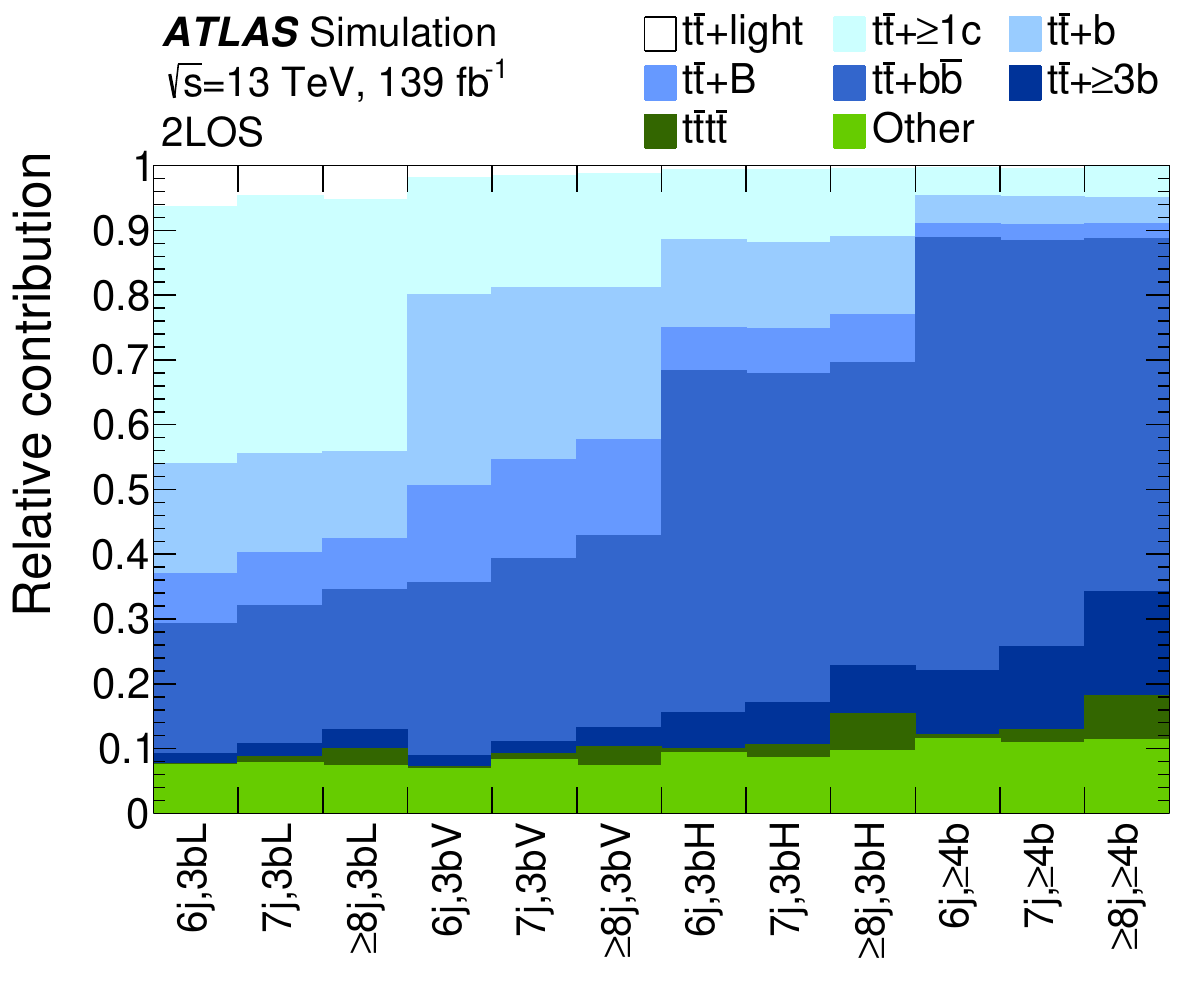}}
\caption{
The relative contribution of the different background classes in the control, validation and signal regions
in the (a) \SL\ and (b) \OS\ channels.
}
\label{fig:cate_frac}
\end{figure}


%
\section{Modelling of \texorpdfstring{\ttjets}{ttbar+jets} background}
\label{sec:ttbar_modelling}

The flavour rescaling factors are determined as corrections to the overall normalisation of
the \ttljets, \ttcjets\ and \ttbjets\ backgrounds.
These factors correct the fraction of the three components
and the overall acceptance of the \ttjets\ background.
To derive the flavour rescaling factors,
events in the 7j and 5j bins respectively in the \SL\ and \OS\ channels are categorised into regions
containing events with two, three or at least four $b$-tagged jets at the 70\% OP\@.
The predictions from the simulations are fit to data simultaneously in these six regions.
The fit observable is the sum of the $b$-tagging pcb scores (defined in Section~\ref{sec:object_preselection})
of the third and fourth jets in descending order of the score.
Systematic uncertainties arising from the calibration of the tagging efficiencies of $b$-jets
and the mis-tag rates of $c$-jets and light-flavour jets are considered~\cite{FTAG-2018-01,FTAG-2020-08,FTAG-2019-02}.
For the \ttljets\ background, two separate rescaling factors are assigned to the \SL\ and the \OS channels.
This is to reduce the tension due to the different acceptance effects of the two channels.
The resulting rescaling factors are $0.84\pm0.04$ ($0.87\pm0.03$) for the \ttljets\ background in the \SL\ (\OS) channel,
$1.61\pm0.13$ for the \ttcjets\ background and $1.18\pm0.03$ for the \ttbjets\ background.
The central values are used to correct the \ttjets\ simulations in the VRs, CRs and SRs.
The fitted uncertainties in the rescaling factors are not considered in the profile likelihood fit determining the signals.
Instead, more conservative uncertainties are assigned
that cover the residual mis-modelling and the extrapolation from the low to high jet multiplicity bins.
For \ttbjets\ and \ttcjets, a normalisation uncertainty of 50\% is assigned to each of them,
based on the discrepancies observed in previous measurements~\cite{ATLAS:2024aht,CMS-TOP-22-009}.
For \ttljets\ events, a normalisation uncertainty of 5\% is considered which covers both the uncertainties in the individual rescaling factors for the \SL\ and \OS\ channels,
and the difference between the two channels.

The NN kinematic corrections are derived from multi-dimensional probability density functions of data and simulations
estimated by using a feed-forward NN~\cite{Garrido_1998,moustakides2019training}.
The NN is trained as a binary classifier between data and simulation.
The output of the NN ($\mathcal{O}$) as a function of the input vector ($\textbf{x}$)
can be interpreted in terms of the a-posteriori Bayesian probability of data and simulations,
$P_\mathrm{data}(\textbf{x})$ and $P_\mathrm{sim}(\textbf{x})$:
\begin{equation}
\mathcal{O}(\textbf{x}) = P(\mathrm{data}|\textbf{x}) = \frac{\alpha_\mathrm{data}P_\mathrm{data}(\textbf{x})}{\alpha_\mathrm{data}P_\mathrm{data}(\textbf{x})+\alpha_\mathrm{sim}P_\mathrm{sim}(\textbf{x})},
\label{eq:nnrew_pdf}
\end{equation}
where $\alpha_\mathrm{data}$ and $\alpha_\mathrm{sim}$ are the proportions of data and simulated events used in the training,
satisfying $\alpha_\mathrm{data} + \alpha_\mathrm{sim} = 1$.
Event-by-event reweighting factors can therefore be obtained as:
\begin{equation}
w(\textbf{x}) = \frac{\alpha_\mathrm{data}P_\mathrm{data}(\textbf{x})}{\alpha_\mathrm{sim}P_\mathrm{sim}(\textbf{x})} = \dfrac{\mathcal{O}(\textbf{x})}{1-\mathcal{O}(\textbf{x})}.
\label{eq:nnrew_w}
\end{equation}
The weights $w(\textbf{x})$ can be used to reweight the simulations to match the probability density function of data.
The training aims to correct only \ttjets\ simulations,
therefore other background events are subtracted from data during the training.
This is done by labelling simulated non-\ttjets\ events as `data' events but with negative event weights.
The events labelled as `simulation' in the training are exclusively \ttjets\ events, corrected with the flavour rescaling factors.
To improve the performance of the training in the regime with small training sample size,
an exponential loss function is used~\cite{moustakides2019training}:
\begin{equation}
\mathcal{L} = P_\mathrm{data}(\textbf{x})e^{-\frac{\mathcal{O}(\textbf{x})}{2}} + P_\mathrm{sim}(\textbf{x})e^{\frac{\mathcal{O}(\textbf{x})}{2}}.
\end{equation}
The resulting reweighting factors after the minimisation of $\mathcal L$ become $w(\textbf{x}) = e^{\mathcal{O}(\textbf{x})}$.
The input variables to the training include the \pt\ of all leptons and jets, the number of jets (\Nj) and large-$R$ jets (\Nrcj) and \met.
The training is performed in the inclusive 2b regions with $\Nj\geq7$ (\SL) and $\Nj\geq5$ (\OS) separately.

Figure~\ref{fig:effect_data_driven_corr} illustrates the effect of the data-driven corrections
in the inclusive region containing all CRs, VRs and SRs,
separately for the \SL\ and \OS\ channels.
The correction improves significantly the agreement between data and the prediction
in both the overall normalisation and the shape of the most relevant kinematic variables for signal-to-background discrimination.

\begin{figure}[]
\centering
\subfloat[]{\includegraphics[width=0.495\textwidth]{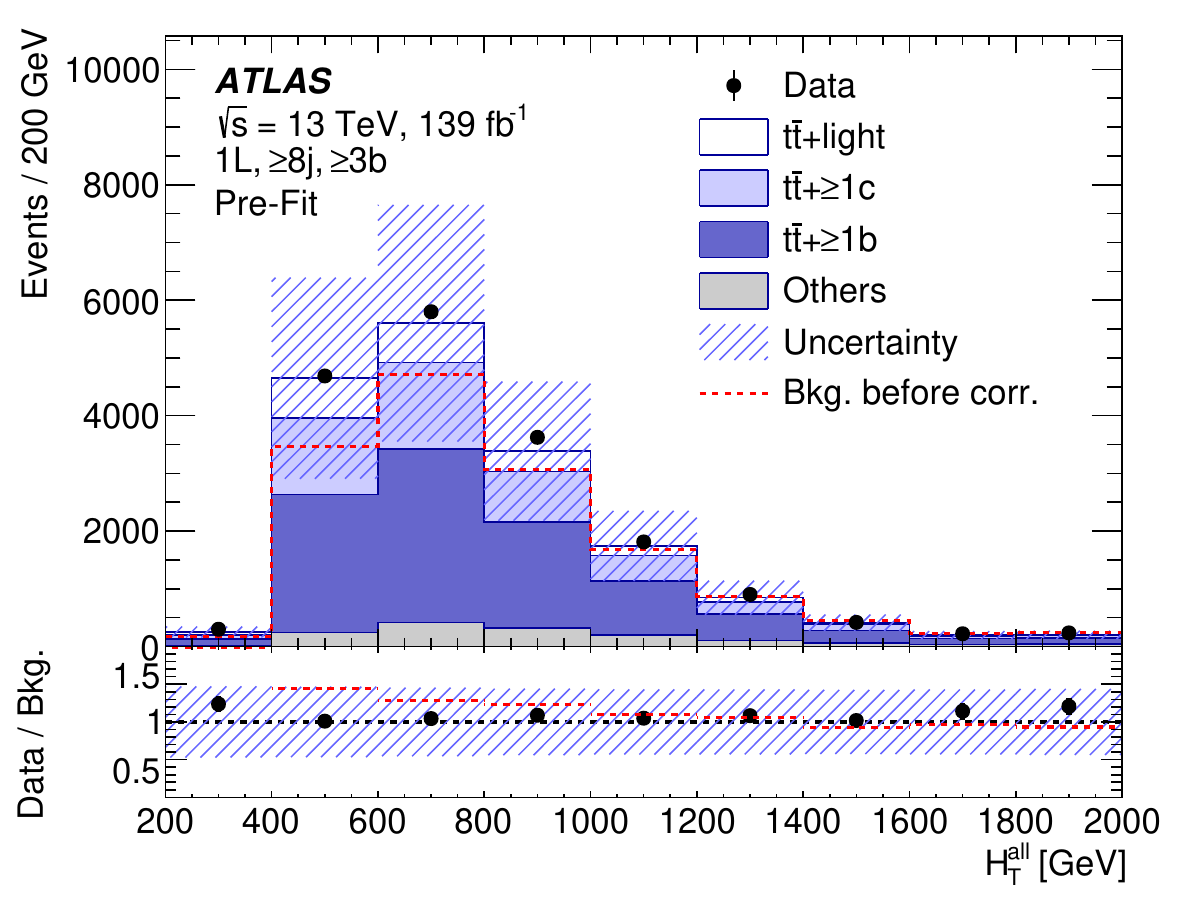}}
\subfloat[]{\includegraphics[width=0.495\textwidth]{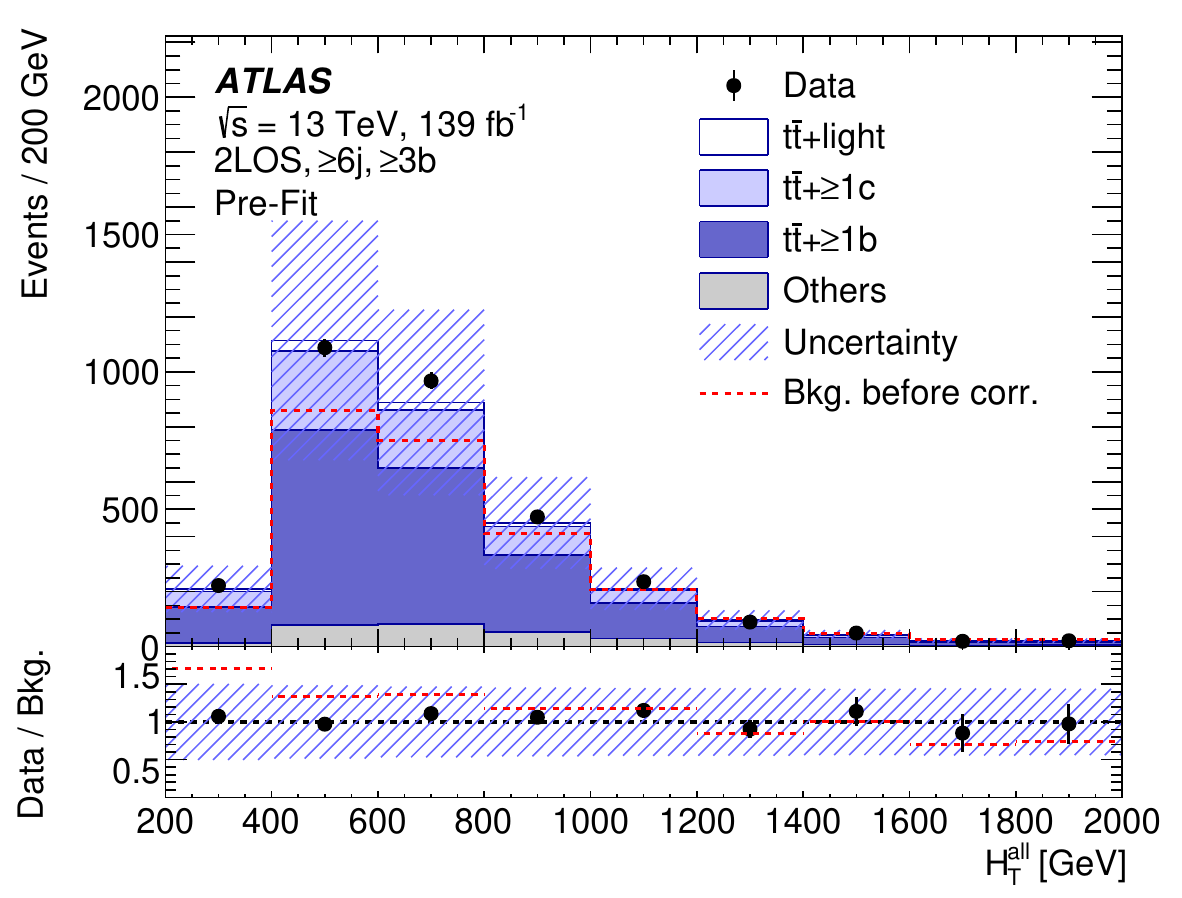}}\\
\subfloat[]{\includegraphics[width=0.495\textwidth]{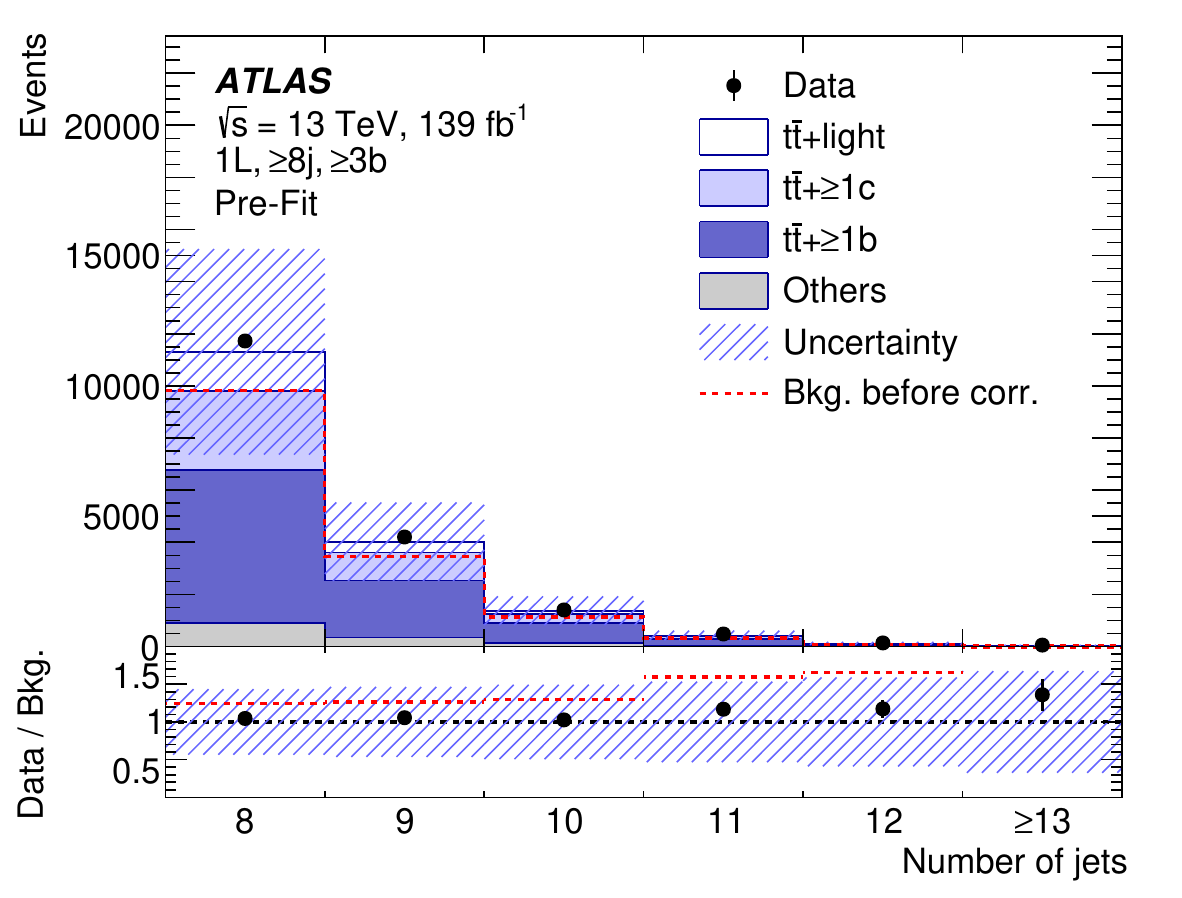}}
\subfloat[]{\includegraphics[width=0.495\textwidth]{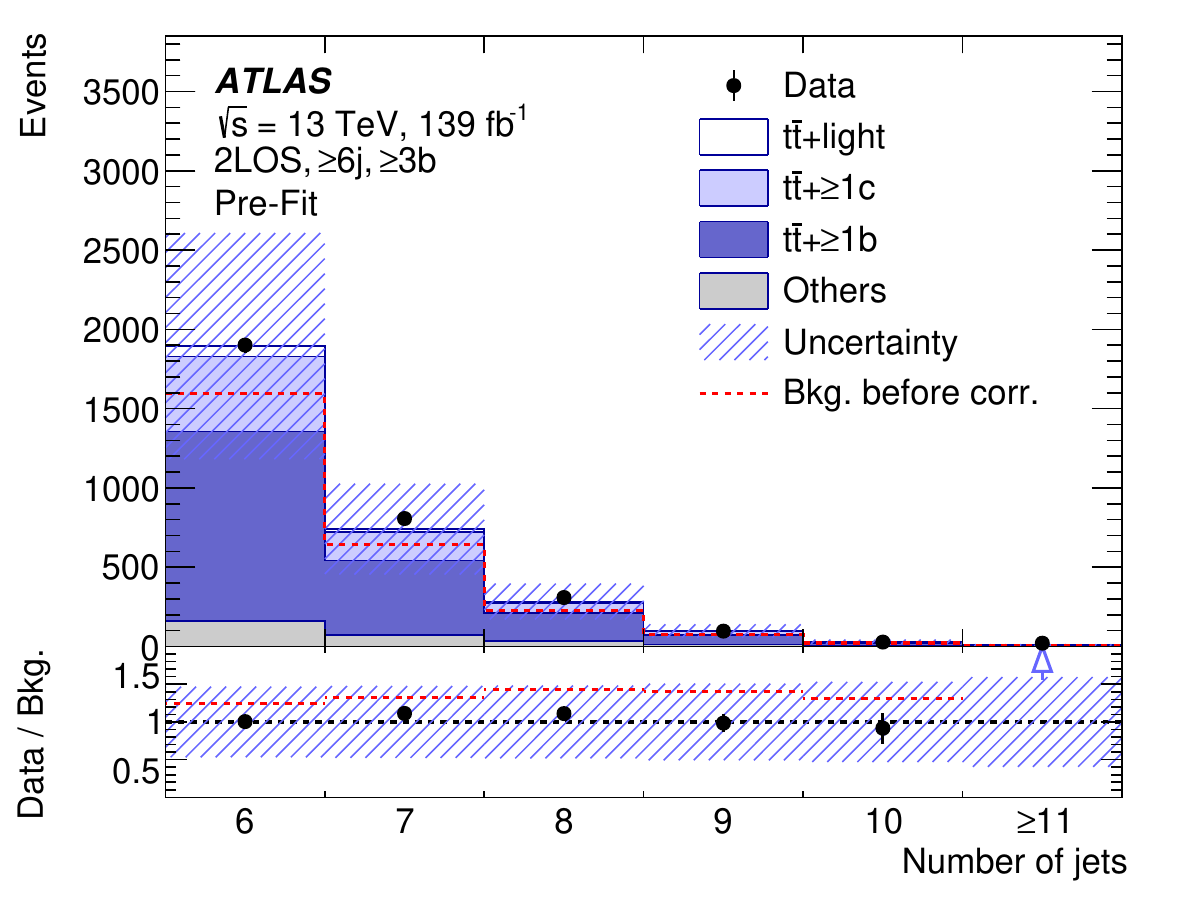}}\\
\subfloat[]{\includegraphics[width=0.495\textwidth]{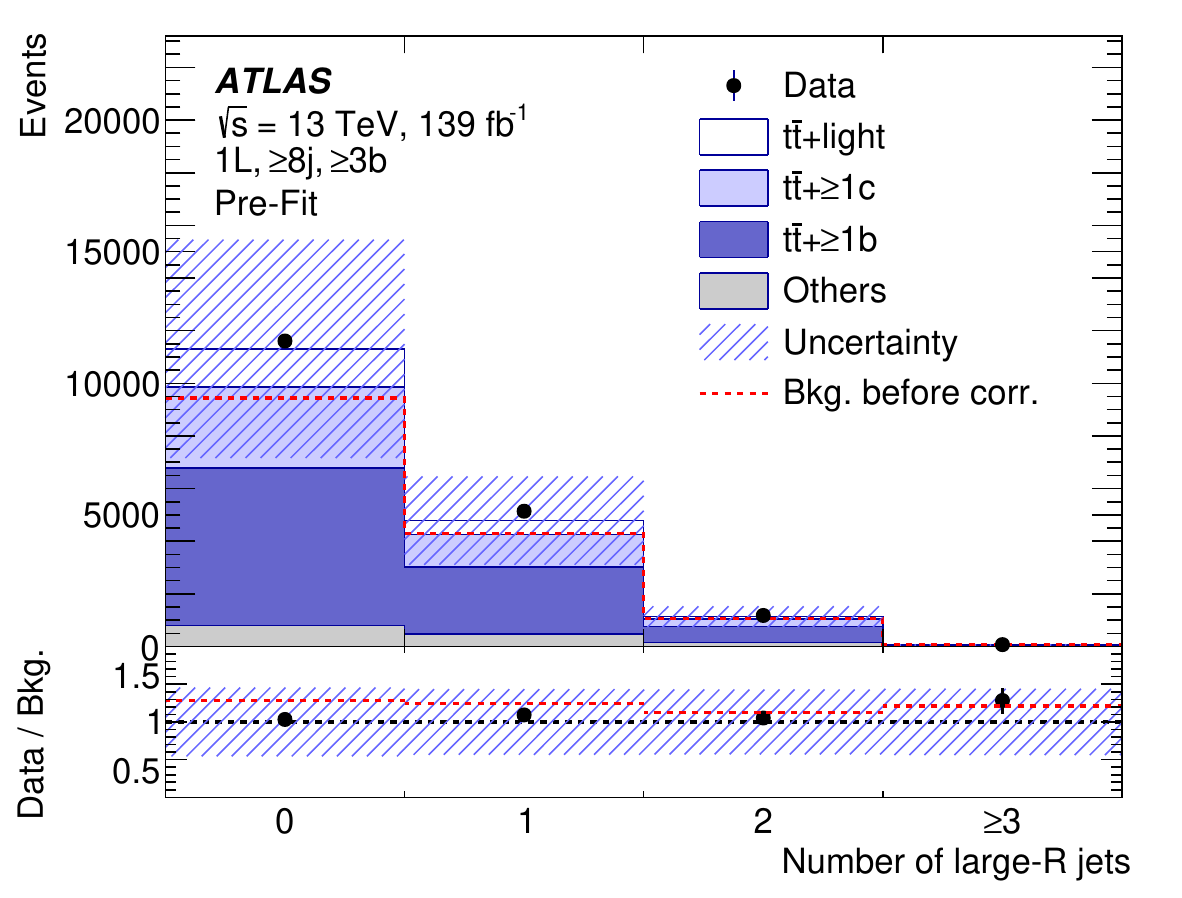}}
\subfloat[]{\includegraphics[width=0.495\textwidth]{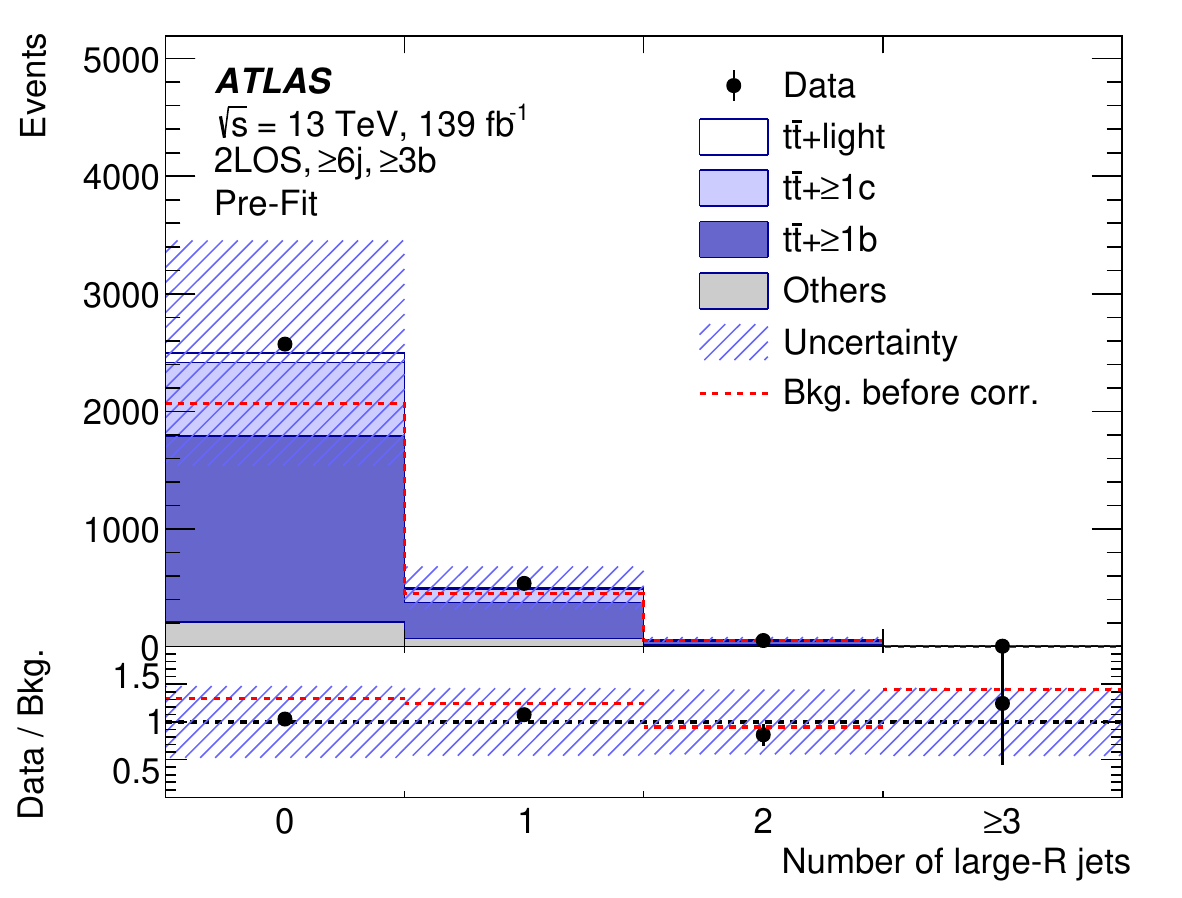}}
\caption{
The comparison of (a, b) \Ht, (c, d) \Nj\ and (e, f) \Nrcj\ distributions between data and background predictions before the likelihood fit to data (Pre-Fit), and before and after applying the data-driven corrections,
separately for the \SL\ and \OS\ channels.
The background prediction after applying the data-driven corrections is shown in coloured stacks,
whilst the one before applying the corrections is shown with the red dashed line. The hashed area represents the total uncertainty in the background. The last bin in each distribution contains the overflow.
}
\label{fig:effect_data_driven_corr}
\end{figure}


%
\section{Multivariate analysis}
\label{sec:mva}

A multivariate discriminant is built based on
a novel GNN structure~\cite{Battgalia_2018} parameterised as $m_{H/A}$~\cite{Baldi_2016}.
The graphs representing the events are constructed using the reconstructed leptons, jets and \met\ as `nodes',
encoding information about their type, four-momenta excluding $\phi$, and jet pcb scores.
`Edges' connecting all pairs of objects in the events encode their relative \dphi, \deta\ and \dR.
In addition, event-level observables are introduced as `global features' of each graph,
covering jet $b$-tagging information, object multiplicities, event shape and kinematics,
and the substructure information of the large-$R$ jets.
These global features are listed in Table~\ref{tab:gnn_global_input}.
The GNNs are built with the \textsc{PyTorch} library~\cite{pytorch_2019}
and trained as a binary classifier with the signal samples at all mass points together against the \ttjets\ background.
The training is performed separately for the \SL\ and \OS\ channels,
using events from the inclusive regions with $\Nj\geq9$ (\SL) and $\Nj\geq6$ (\OS) and $N_b^{70\%}\geq3$.
To balance the resulting performance against different signal masses,
the signal samples at the different mass points are normalised to the same yields in the defined inclusive regions.
The output of the GNN is used as the discriminant in the SRs.
The 3bV regions with $\geq$9 ($\geq$7) jets in the \SL\ (\OS) channel
are used to validate the modelling of the GNN output.
The $m_{H/A}$ parameterisation is done by including $m_{H/A}$ as an input to the training alongside the graphs.
For the signal samples, this corresponds to the generated $m_{H/A}$.
In the \ttjets\ samples, a pseudo-$m_{H/A}$ is assigned to each event
randomly taking the values from the generated signal $m_{H/A}$ with a uniform prior.
The parameterisation allows for a smooth interpolation between the results at different $H/A$ mass points~\cite{Baldi_2016}.

\begin{table}[t!]
\centering
\begin{center}
\caption{
The list of global features used in the GNN training
in descending order of importance in the training of the \SL\ GNN\@.
The importance is evaluated according to the fractional change in loss when removing the feature from the training.
}
\label{tab:gnn_global_input}
\renewcommand{\arraystretch}{1.2}
\begin{tabular}{ll}
\toprule \midrule
Variable                      &    Description \\
\midrule
$\sum_{i\in[1,6]}\mathrm{pcb}_i$    & Sum of the pcb scores of the six jets with the highest scores \\
\Ht                           & \pt\ sum of all reconstructed leptons and jets \\
\Nj                           & Number of jets \\
$\Ht^\mathrm{ratio}$          & \pt\ sum of the four leading jets in \pt\ divided by the \pt\ sum of the remaining jets \\
$dR_{jj}^\mathrm{avg.}$       & Average \dR\ across all jet pairs \\
$m_\mathrm{T}^W$              & $W$ boson transverse mass calculated using the lepton four-momenta and \met (\SL\ only) \\
$\dR_{bb}^\mathrm{min.}$      & Minimum \dR between any pair of jets $b$-tagged at the 70\% OP \\
$\dR_{\ell b}^\mathrm{min.}$  & Minimum \dR between any pair of lepton and jet $b$-tagged at the 70\% OP \\
$m_{bbb}^\mathrm{avg.}$       & Average invariant mass of all triplets of jets $b$-tagged at the 70\% OP \\
$m_{jjj}^\mathrm{avg.}$       & Average invariant mass of all triplets of jets with an angular separation of $\dR<3$ \\
$\sum d_{12}$                 & Sum of the first $k_t$ splitting scale $d_{12}$ over all large-$R$ jets \\
$\sum d_{23}$                 & Sum of the second $k_t$ splitting scale $d_{23}$ over all large-$R$ jets \\
\Nrcj                         & Number of large-$R$ jets with a mass greater than 100~\GeV \\
Centrality                    & $\sum_i \pt^i/\sum_i E_i$ where the sums are performed over all reconstructed jets and leptons \\
$m_{\ell\ell}$                & Invariant mass of the two leptons (\OS\ only) \\
\midrule \bottomrule
\end{tabular}
\end{center}
\end{table}

The training achieves an area under the receiver-operator-characteristic curve of
85\% for the 400~\GeV\ mass point and 94\% for 1000~\GeV\ mass point.
The dominant feature for the training is the node-\pt\ for both the channels,
followed by $\sum_{i\in[1,6]}\mathrm{pcb}_i$ and the pcb scores of the jet-nodes.
Collectively, the global features \Ht, \Nj\ and $\Ht^\mathrm{ratio}$
and the edge feature \dR\ also contribute significantly.
The distributions of the GNN output evaluated at 400~\GeV\ and 1000~\GeV\
for both the \SL\ and \OS\ channels are shown in Figure~\ref{fig:mva_separation_m400}.
The signal is compared with the SM \tttt\ and \ttjets\ backgrounds.
The distributions of the signal and SM \tttt\ production is similar for $m_H = 400$~\GeV.
The discrimination is significantly better for large $m_H$.

\begin{figure}[ht!]
\centering
\subfloat[]{\includegraphics[width=0.45\textwidth]{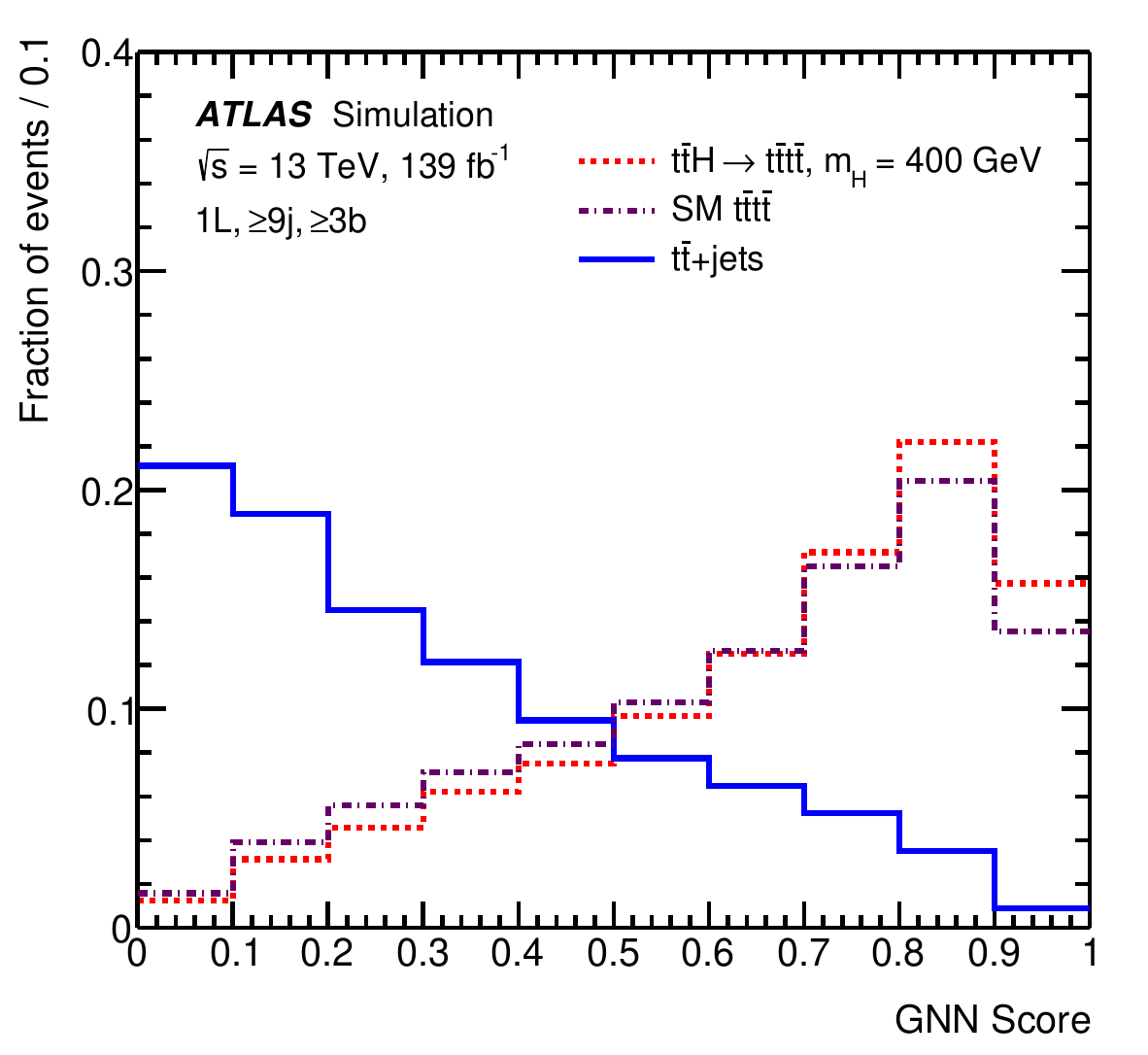}}
\subfloat[]{\includegraphics[width=0.45\textwidth]{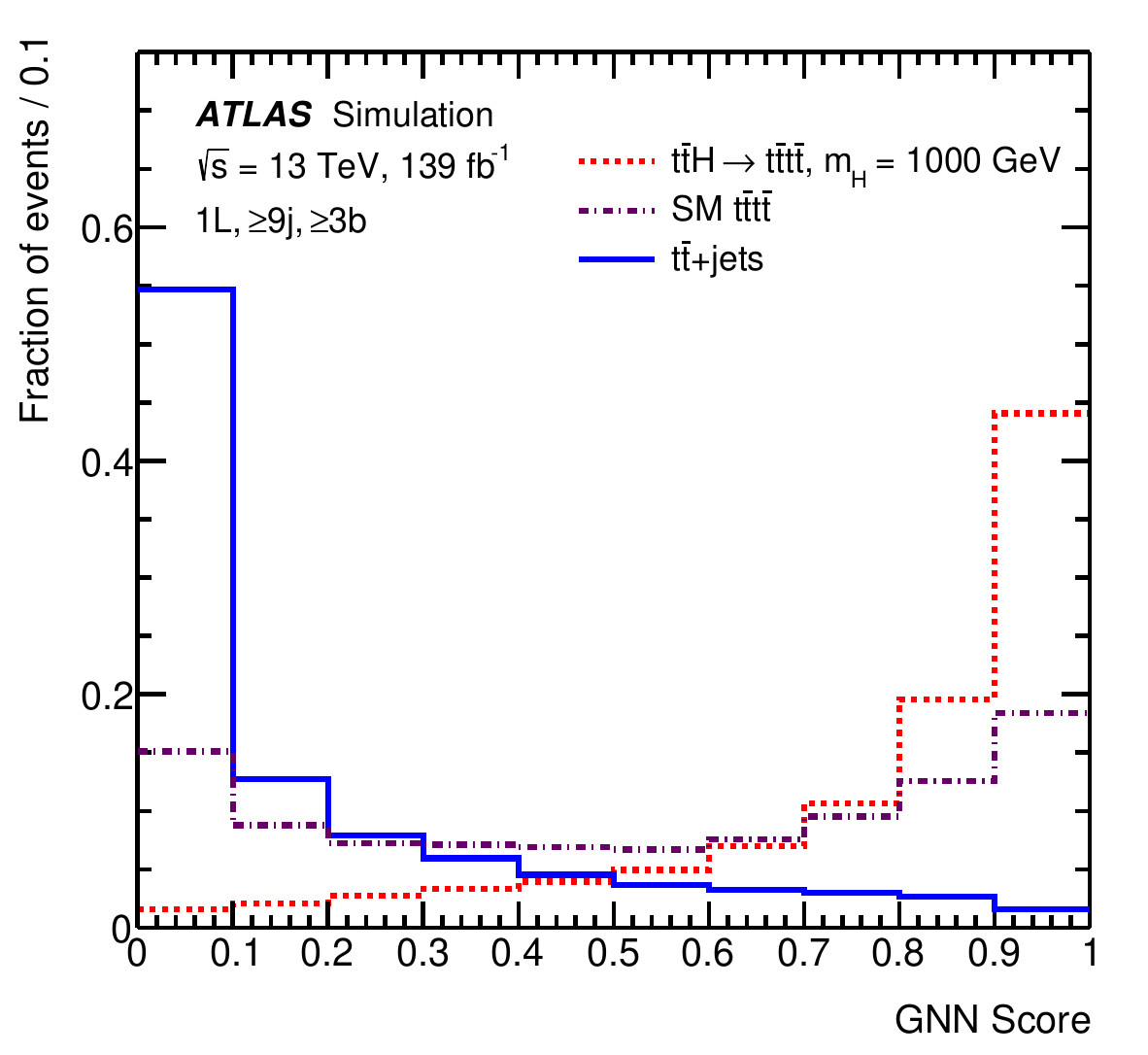}}\\
\subfloat[]{\includegraphics[width=0.45\textwidth]{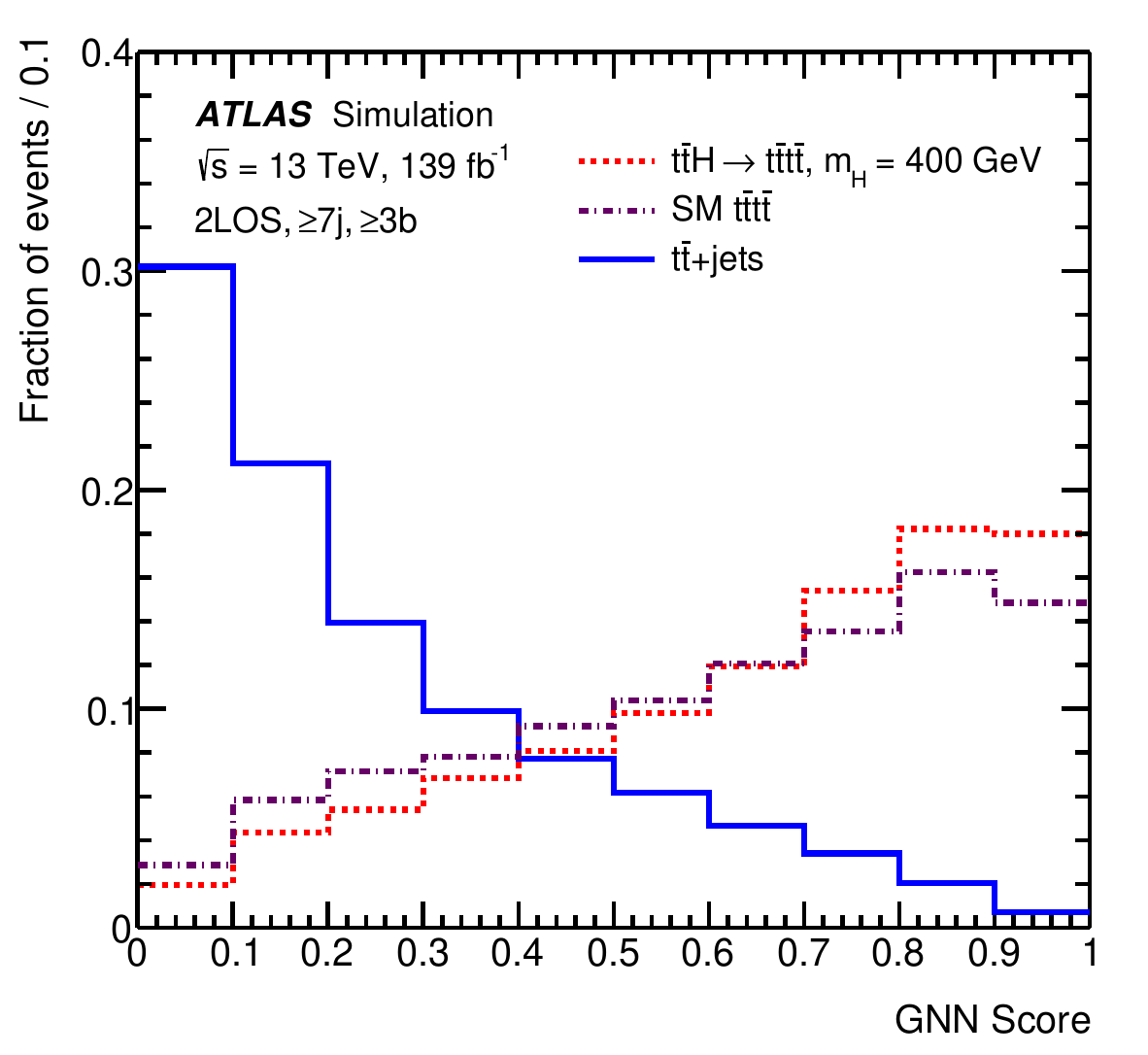}}
\subfloat[]{\includegraphics[width=0.45\textwidth]{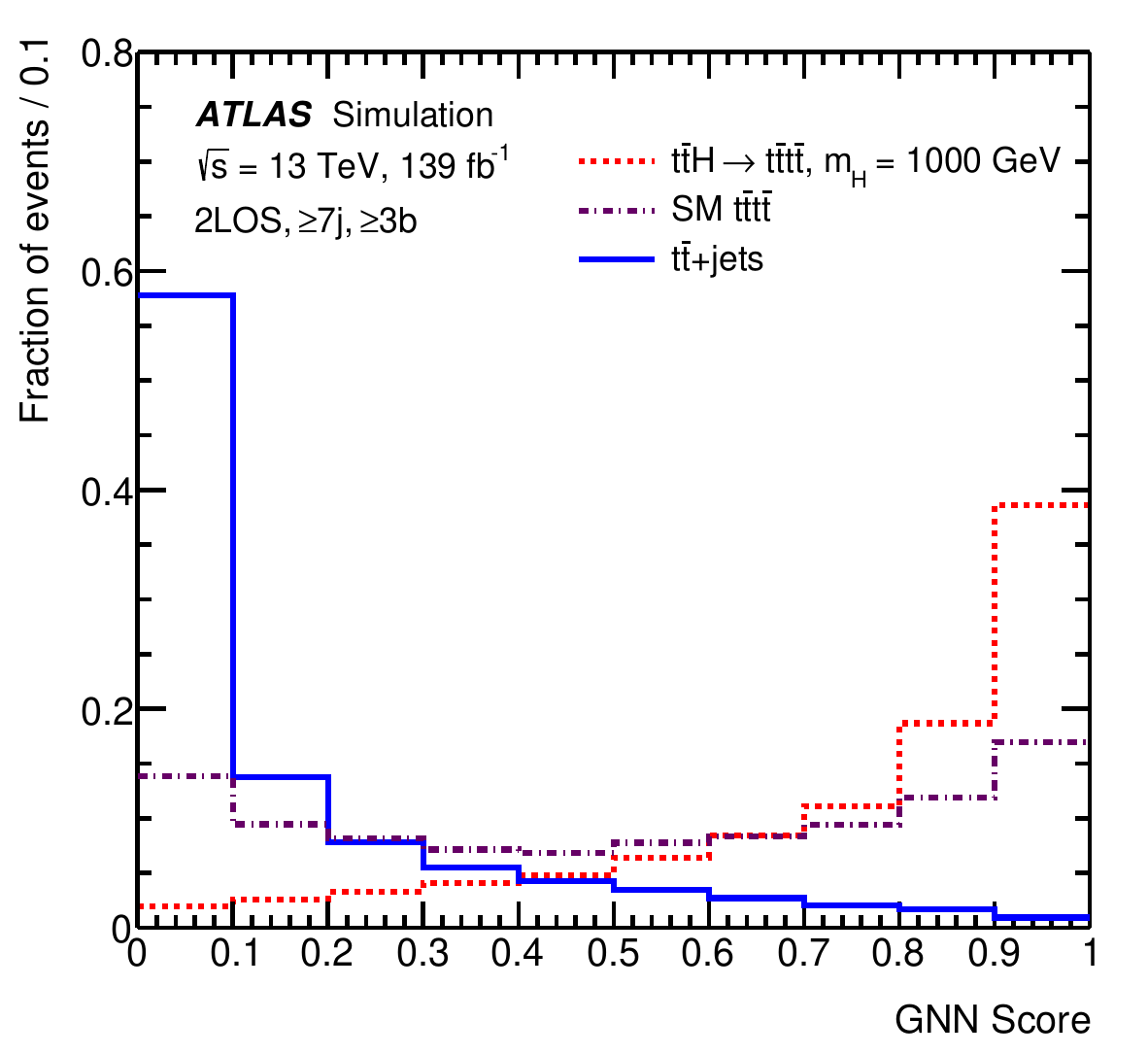}}
\caption{
The distributions of the GNN output evaluated at 400~\GeV\ and 1000~\GeV\ for the (a, b) \SL\ and (c, d) \OS\ channels with inclusive selections close to the signal region.
The distributions are compared for the signal, SM \tttt\ production and \ttjets\ production.
}
\label{fig:mva_separation_m400}
\end{figure}


%
\section{Systematic uncertainties}
\label{sec:systematics}
Systematic uncertainties in the results arise from uncertainties in
experimental procedures, signal and background cross-sections,
and the modelling of the relevant physics processes.
Each uncertainty can affect either or both the normalisation and shape of the predictions.
The largest uncertainties are from the modelling of the \ttjets\ background,
especially  \ttbjets\ events.
The uncertainties in the modelling of the SM \tttt\ production also have a sizeable contribution.
The impact of the experimental uncertainties is small.
The effects of the post-fit uncertainties on the results are summarised in Table~\ref{tab:grouped_impact} in Section~\ref{sec:results}.

\subsection{Experimental uncertainties}

All of the  individual experimental uncertainties have a small impact on the analysis.
The leading group of experimental uncertainties arises from
the calibration of the jet energy scale and resolution~\cite{JETM-2018-05}.
The most important components are related to
the modelling of the composition of gluon- and quark-jets and their detector response.
The calibration of the $b$-tagging efficiencies and $c$- and light-jet rejection rates
have smaller effects amongst the experimental uncertainties~\cite{FTAG-2018-01,FTAG-2020-08,FTAG-2019-02}.
Other experimental uncertainties have minor contributions.
The uncertainty in the integrated luminosity of the  data sample is 1.7\%~\cite{ATLAS-CONF-2019-021},
measured using the LUCID-2 detector~\cite{LUCID2}
with complementary measurements using the ID and calorimeters.
An uncertainty in the correction of the pile-up profile in the simulations to match that in data is considered.
An uncertainty is applied to the corrections on the JVT selection efficiencies in simulations~\cite{PERF-2014-03}.
Uncertainties in electrons and muons arise from
the calibration of the efficiencies of the trigger, reconstruction, identification and isolation requirements,
and the energy scale and resolution~\cite{JETM-2020-03}.
Uncertainties in the measurement of \met\ arise from a possible mis-calibration of its soft-track component~\cite{PERF-2016-07}.

\subsection{Uncertainties in the signals, SM \texorpdfstring{\tttt}{tttt} and \texorpdfstring{\ttjets}{tt+jets}}
\label{subsec:ttbar_unc}

The uncertainties in signal modelling have minor impacts on the results.
The uncertainties due to missing higher order QCD corrections and from PDF variations are considered.
The former is evaluated by varying the renormalisation and factorisation scales in \MGNLO\
by a factor of two up and down, and comparing to the nominal prediction.
The latter has an effect of up to 1\% on the signal yields, estimated following the PDF4LHC recommendations~\cite{Butterworth:2015oua}.

The largest uncertainty in \tttt\ modelling
is evaluated by comparing the nominal sample simulated using \MGNLO+ \PYTHIA[8] with the one using \SHERPA[2.2.11].
This uncertainty covers the effects from the different matching schemes between the ME and PS,
and  different PS, hadronisation and fragmentation models.
The uncertainty in the predicted \tttt\ production cross-section,
$^{+7.8\%}_{-13.3\%}$ from Ref.~\cite{vanBeekveld:2022hty},
has a smaller impact.
Other uncertainties in \tttt\ modelling have negligible impacts.
These include
an uncertainty evaluated by comparing the nominal sample with the one simulated using \MGNLO+ \HERWIG[7],
and the uncertainties from missing higher order QCD corrections and PDF variations
evaluated in the same way as those for the signals.

The uncertainties in \ttjets\ modelling are decomposed based on the \ttjets\ event classification.
The uncertainties for the different event classes are treated as independent.
Uncertainties in the overall normalisation of \ttljets\ (5\%), \ttcjets\ (50\%) and \ttbjets\ (50\%)
are assigned as described in Section~\ref{sec:ttbar_modelling}.
Three sets of uncertainties are evaluated by comparing samples simulated using different generators.
The effects from the ME-PS matching and the PS, hadronisation and fragmentation models
are covered by the two uncertainties from the comparison of the nominal \POWHEG+ \PYTHIA[8] sample
with those simulated using \MGNLO+ \PYTHIA[8] and \POWHEG+ \HERWIG[7].
For \ttbjets\ events, another uncertainty is assigned
by comparing the nominal prediction based on the 5FS
with the one from the 4FS $\ttbar b\bar{b}$ sample.
Each of these three uncertainties is first decomposed into components based on the finer classification that considers \ttbjets\ subclasses.
The component for each event class is further split into two,
one controlling the shape of the fitted distributions in each region
and another controlling the relative normalisation across the different regions.
Other uncertainties are evaluated by varying the value of physics parameters used in the generation of the nominal sample.
These are split into components for \ttljets, \ttcjets\ and \ttbjets.
The uncertainty from missing higher order QCD corrections
is estimated by varying the factorisation and renormalisation scales
individually or simultaneously by a factor of two up or down,
and taking the largest up and down variations in any of the analysis bins.
An uncertainty due to the chosen \hdamp\ value is estimated by varying its value up by a factor of two.
The uncertainties in the amount of initial-state and final-state radiations (ISR/FSR) from the PS
are evaluated, respectively,
by varying the A14 tune parameter that controls the strong coupling
and the FSR renormalisation scale by a factor of 2.0 and 0.625.
The PDF uncertainties are estimated by using the same method as for the signals.

\subsubsection{Flavour rescaling and NN kinematic corrections on \texorpdfstring{\ttjets}{tt+jets} modelling uncertainties}

The data-driven corrections described in Section~\ref{sec:ttbar_modelling}
are consistently derived for and applied to the alternative \ttjets\ predictions used to evaluate modelling uncertainties (Section~\ref{subsec:ttbar_unc}).
For each of these alternative predictions,
dedicated flavour rescaling factors and NN kinematic corrections are derived.
The flavour rescaling factors are derived such that the alternative predictions have the same \ttbjets, \ttcjets\ and \ttljets\ yields as the nominal prediction after the flavour rescaling.
The NN kinematic corrections are obtained from dedicated GNN trainings
where the corresponding alternative \ttjets\ simulations are used.
These corrections reduce the size of the pre-fit uncertainties by exploiting the data.
Evaluating dedicated flavour rescalings for each modelling uncertainty
also avoids double counting effects from the acceptance and \ttHF\ production cross-section,
significantly reducing the correlations between these
uncertainties and the \ttbjets, \ttcjets\ and \ttljets\ normalisation uncertainties.
For experimental uncertainties, given their small impact on the predictions, the nominal rescaling factors and NN kinematic corrections are used.

\subsubsection{Uncertainties in NN kinematic corrections}

Two sets of uncertainties are considered for the derived NN kinematic corrections.
To account for the uncertainties in the cross-sections of the subtracted non-\ttjets\ background processes,
their cross-sections are varied by $\pm$50\% simultaneously and the NN is retrained.
The resulting uncertainties have a 2\% to 6\% effect on the fitted distributions.
In addition, the influence of the finite training data sample is evaluated
using the MC dropout method~\cite{gal2016dropout}.
The NN is retrained 1000 times, each time with randomly dropped-out neurons.
The standard deviation of the NN outputs from the 1000 sets of training output
are propagated to the fitted distributions.
The resulting uncertainties range from about 5\% in the CRs up to 12\% in the SRs.

\subsection{Other uncertainties}

Other small uncertainties arise from the modelling of the minor backgrounds,
and the uncertainties due to the finite size of the simulated samples,
referred to as `simulation statistical uncertainties'.
The latter are treated as independent for each analysis bin,
and separately for the total background and any particular signal sample considered in the fit.
The cross-section uncertainties for the minor background are assigned taking into account
the high jet multiplicity regime.
For \ttW, \ttZ\ and \ttH\ production,
uncertainties of 60\%~\cite{TOPQ-2018-05}, 15\%~\cite{deFlorian:2016spz} and 20\%~\cite{HIGG-2018-57} are assigned to
the inclusive production cross-sections.
Additional uncertainties of 10\%, 20\% and 30\% are assigned to
$\ttbar W/Z/H$ events with nine (seven), ten (eight), $\geq$11 ($\geq$9) jets in the \SL\ (\OS) channel,
based on the discrepancies observed between data and \ttjets\ simulations in the 2b regions
before applying the data-driven corrections.
For the \Vjets\ backgrounds, an uncertainty of 60\% is assigned,
estimated by adding in quadrature a 24\% uncertainty for each additional jet
based on a comparison of different algorithms for merging LO ME and PS generators~\cite{Alwall:2008epjc}.
Uncertainties of 30\% and 35\% are assigned to the single-top-quark and \ttt\ production
cross-sections~\cite{TOPQ-2021-08}.
For $\ttW/Z/H$ and single-top-quark production,
uncertainties due to missing higher-order QCD corrections are estimated by using the same method used for the signals.
The alternative samples described in Section~\ref{sec:mc} are compared with the nominal samples
to evaluate effects from the ME-PS matching or the modelling of the PS, hadronisation and fragmentation.
For single-top-quark production, an additional uncertainty is evaluated
by comparing the nominal prediction with the one simulated using the diagram subtraction scheme.


%
\section{Results and interpretation}
\label{sec:results}

\subsection{Statistical analysis}
\label{sub:stat_analysis}

The \Ht\ distributions in the CRs and the GNN output distributions in the SRs
are used to search for the presence of a signal.
The simulation is fit to data,
using a profile likelihood function ${\cal L}(\mu,\pmb{\theta})$ that is constructed as
a product of Poisson and Gaussian probability density terms over all bins in the distributions of the CRs and SRs:
\begin{eqnarray}{\cal L}\left(\pmb{n}|\mu, \pmb{\theta}\right) = \prod_{r \in \text{region}}\prod_{i\in \text{bin}} \text{Pois}\left(n_{i,r}|\mu S_{i,r}(\pmb{\theta}) + B_{i\
,r}(\pmb{\theta})\right)\times\prod_{j\in \text{NP}}G\left(\theta_j\right).
\label{eq:Likelihood}
\end{eqnarray}
Here $\mathbf{n}$ represents the data vector, with $n_{i,r}$ representing data yields in the bin $i$ and
region $r$. The vector $\pmb{\theta}$ denotes the nuisance parameters (NPs), each representing a systematic uncertainty,
that affect the number of signal events $S_{i,r}$ as well
as the number of background events $B_{i,r}$ in the bin $i$ and region $r$. The quantity $\mu$ is the signal strength,
which is the parameter of interest in the likelihood and it represents the ratio of the measured cross-section
to a reference cross-section used to normalise the signal samples.
This cross-section is arbitrary and does not change the cross-section upper limit.
The terms `$\mathrm{Pois}$' and `$\mathrm{G}$' represent the Poisson and Gaussian probability distributions respectively.
The Gaussian distribution terms adjust the expectations for signal and background according to their systematic uncertainties.

For limit setting, a test statistic  $q(\mu)$ is defined for use with the CL$_{s}$ method~\cite{Junk:1999kv,Read:2002hq}.
The profile likelihood ratio used is:
$q(\mu) = -2\ln({\cal L}(\mu,\hat{\hat{\pmb{\theta}}}_\mu)/{\cal L}(\hat{\mu},\hat{\pmb{\theta}}))$,
where $\hat{\mu}$ and $\hat{\pmb{\theta}}$ are the values of the parameters that
maximise the likelihood function (with the constraint $0\leq \hat{\mu} \leq \mu$),
and $\hat{\hat{\pmb{\theta}}}_\mu$ are the values of the
NPs that maximise the likelihood function for a given value of $\mu$.
No significant excess above the background expectation is observed and therefore upper limits on $\mu$ for each of the
signal scenarios considered are derived.
For a given signal scenario, values of $\mu$ yielding CL$_{s}$<0.05,
where CL$_{s}$ is computed using the asymptotic  approximation~\cite{Cowan:2010js}, are excluded at $\geq$95\% confidence level (CL).
All the statistical analyses have used the \textsc{RooStats} framework~\cite{Cranmer:1456844,RooFit,moneta2011roostats}.

For each $m_{H}$ scenario,
a profile likelihood fit is performed using all CRs and SRs in both the \SL\ and \OS\ channels simultaneously.
The post-fit distributions of the GNN output for the most sensitive regions in the \SL\ and \OS\ channels
are shown in Figure~\ref{fig:post_fit}.
The GNN scores are evaluated with $m_{H/A} = 400$~\GeV.
The same distributions in the VRs that are the closest to the SRs in the \SL\ and \OS\ channels are shown in Figure~\ref{fig:post_fit_vr},
illustrating a good agreement between data and post-fit background prediction.

\begin{figure}[t!]
\centering
\subfloat[]{\includegraphics[width=0.495\textwidth]{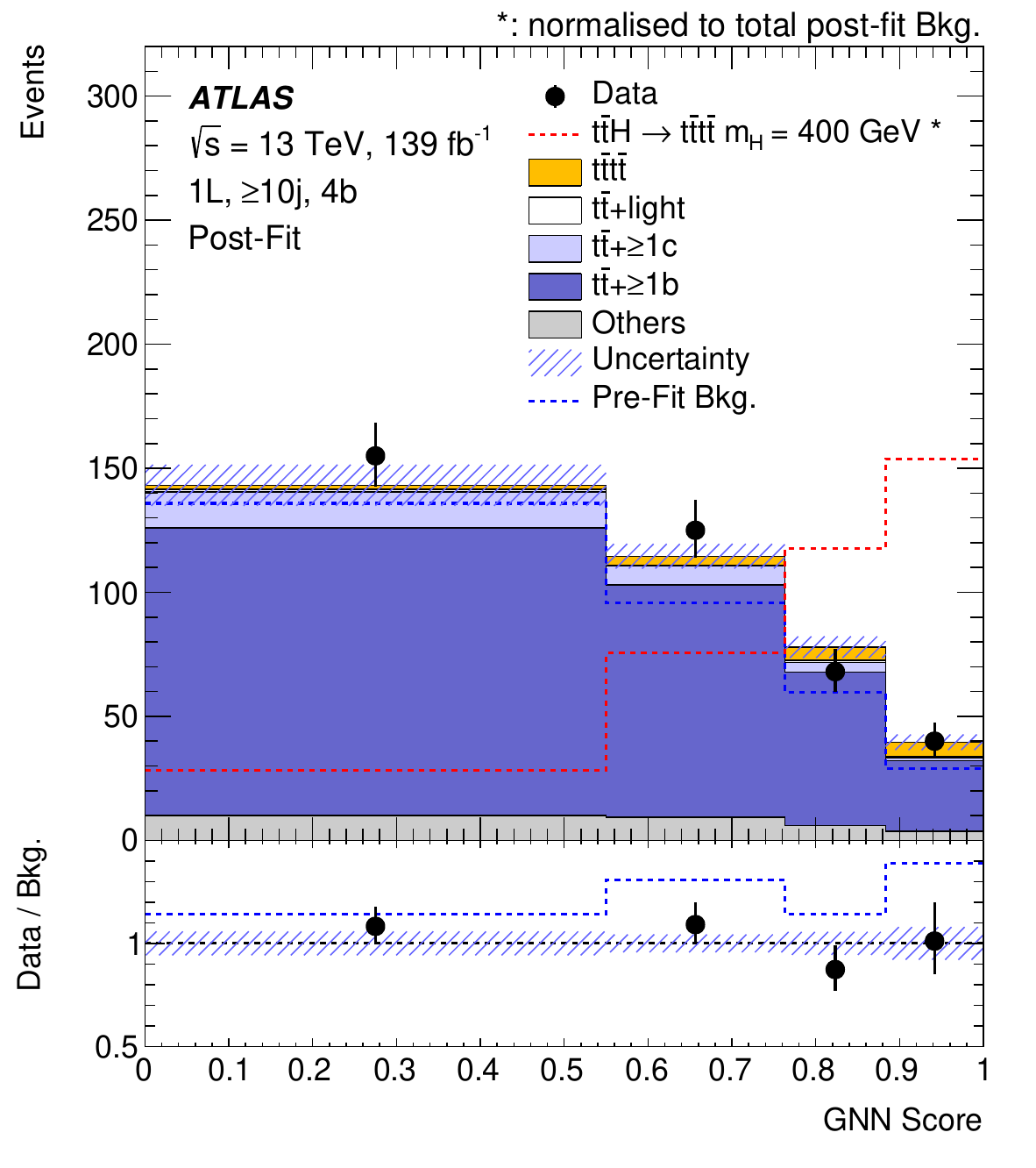}}
\subfloat[]{\includegraphics[width=0.495\textwidth]{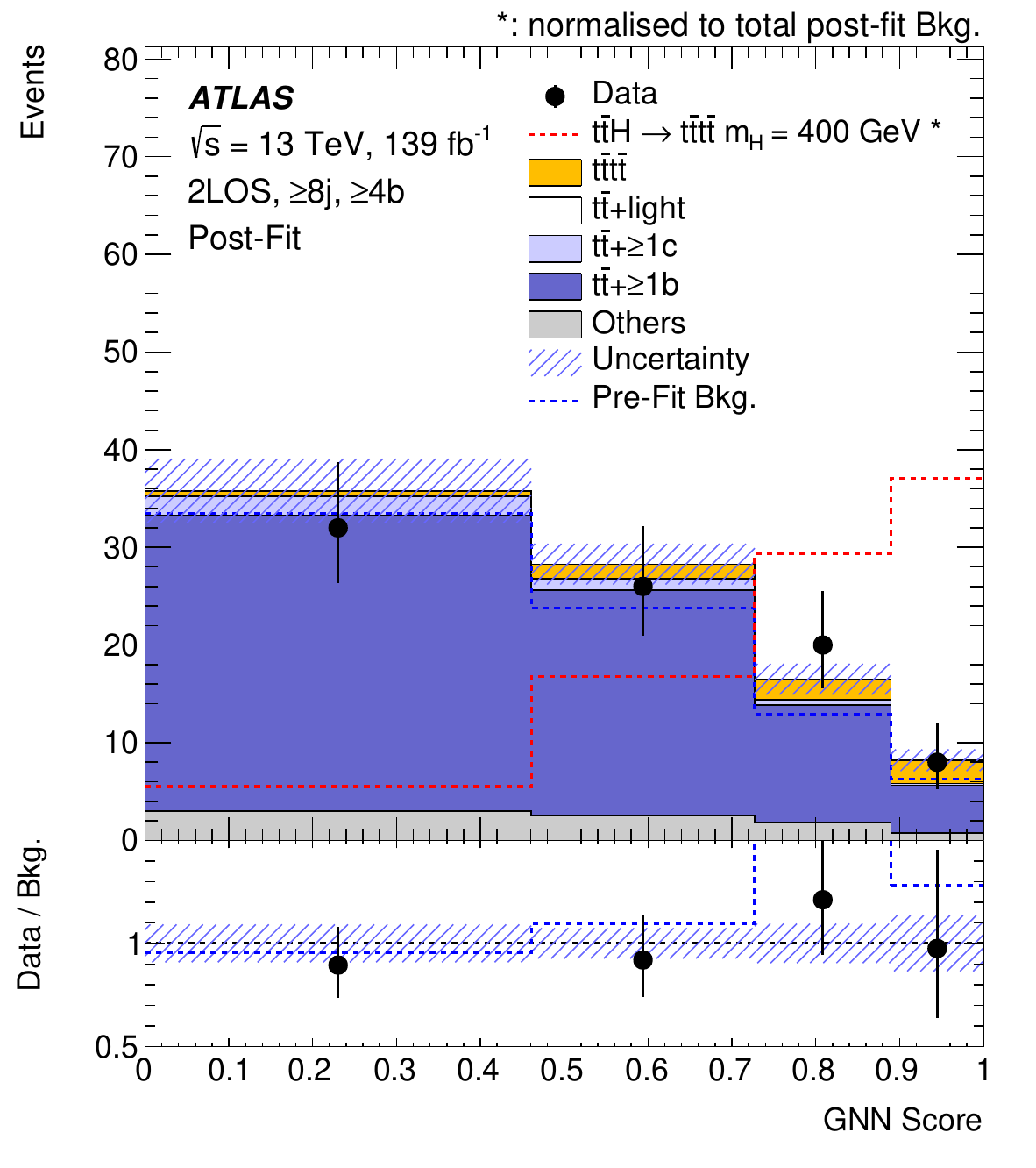}}
\caption{Post-fit distribution of the GNN score evaluated with $m_{H/A} = 400$~\GeV\
(a) in the \SL\ region with $\geq 10$ jets and four $b$-tagged jets and
(b) in the \OS\ region with $\geq 8$ jets and $\geq 4$ $b$-tagged jets.
The fit is performed under the background-only hypothesis.  The hashed area
represents the total uncertainty in the background.
The distribution of the signal (red dashed line) is shown normalised to the total post-fit background for illustration purposes.
The pre-fit background prediction is illustrated using a blue dashed line.
}
\label{fig:post_fit}
\end{figure}

\begin{figure}[t!]
\centering
\subfloat[]{\includegraphics[width=0.495\textwidth]{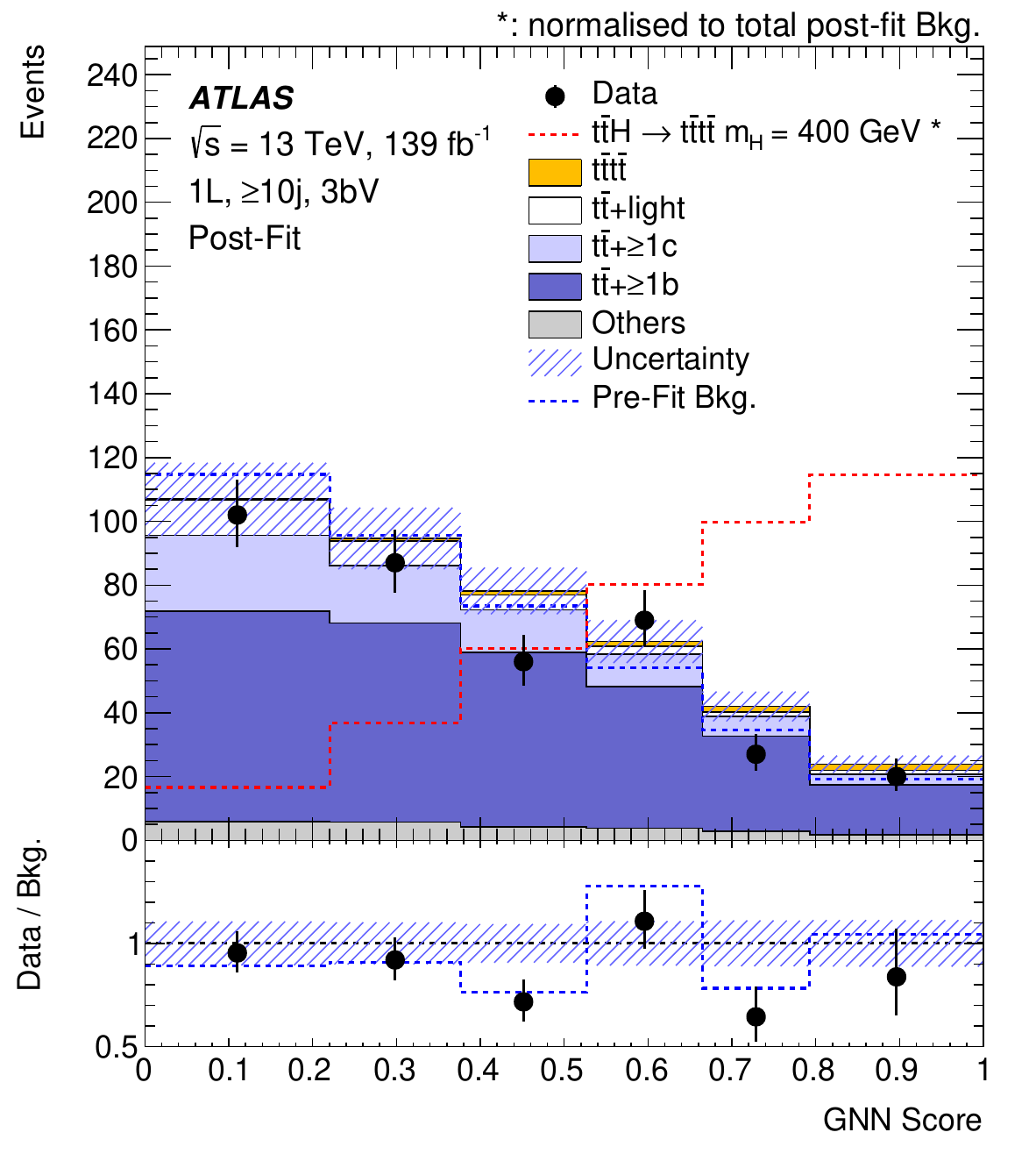}}
\subfloat[]{\includegraphics[width=0.495\textwidth]{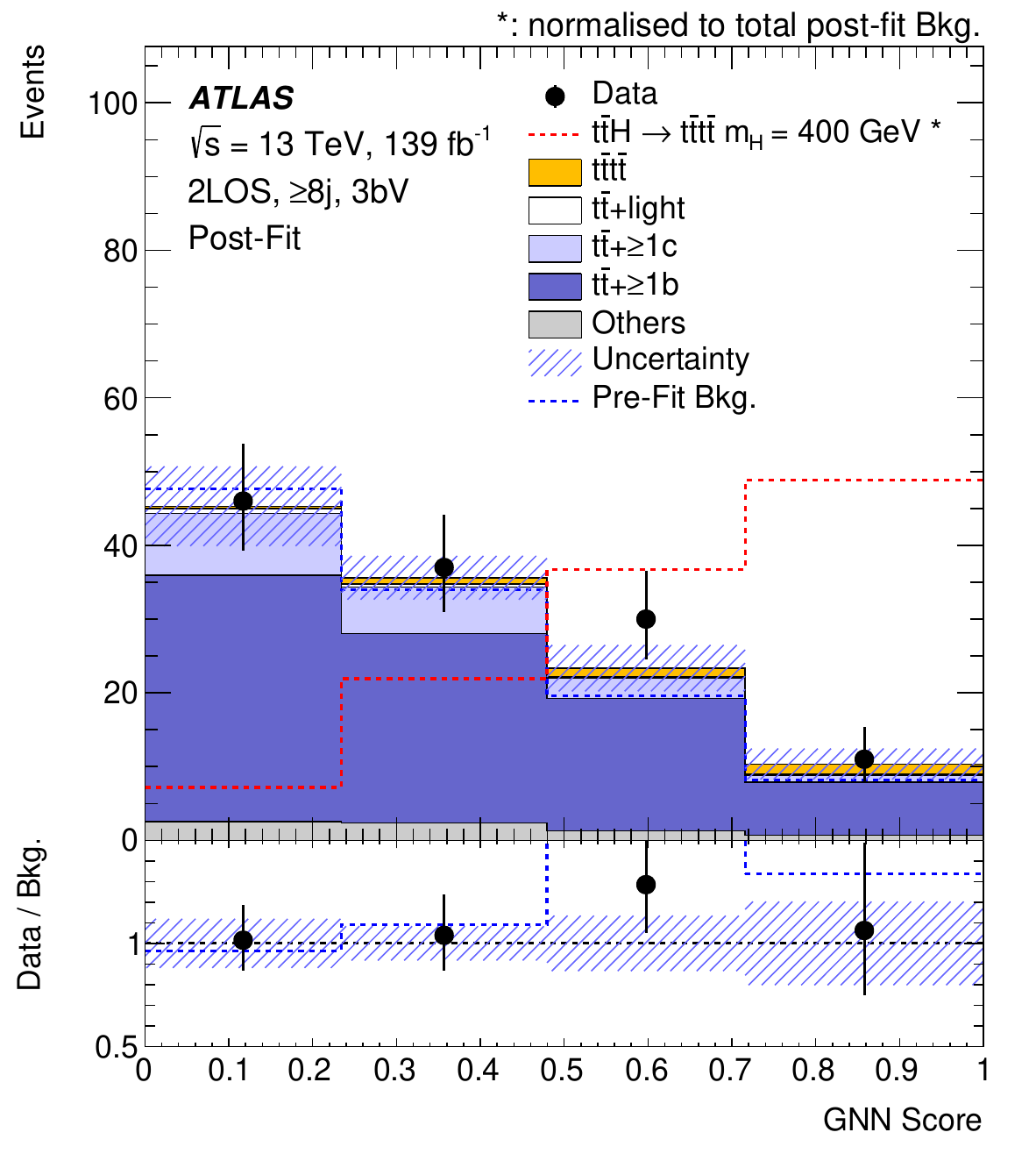}}
\caption{Post-fit distribution of the GNN score evaluated with $m_{H/A} = 400$~\GeV\ in the VRs
(a) in the \SL\ region with $\geq 10$ jets and
(b) in the \OS\ region with $\geq 8$ jets.
These regions do not enter the fit.
The post-fit background prediction is obtained using the post-fit nuisance parameters
from the background-only fit in the CRs and SRs.
The hashed area represents the total uncertainty in the background.
The distribution of the signal (red dashed line) is shown normalised to the total post-fit background for illustration purposes.
The pre-fit background prediction is illustrated using a blue dashed line.
}
\label{fig:post_fit_vr}
\end{figure}

The search in the \SLOS\ final states is combined with the previous analysis in the \SSML\ final states~\cite{EXOT-2019-26}.
The combination is performed via a simultaneous profile likelihood fit including all SRs and CRs of both the final states,
with all systematic uncertainties included.
The events in the two analyses are statistically independent given the lepton selection criteria.
The experimental uncertainties are treated as fully correlated between the two analyses since both the
analyses use the same reconstructed objects and data set.
Theoretical modelling uncertainties in the non-\ttjets\ backgrounds and the signals are also
correlated if they have identical definition in both the \SLOS\ and \SSML\ final states.
The \ttW\ and \ttjets\ backgrounds are treated differently in the two analyses.
The \ttjets\ background is the
dominant irreducible background in the \SLOS\ final state and data-based corrections are applied.
In the \SSML\ final state this background is a much smaller contribution
and arises from prompt leptons with mis-identified charges
or from mis-reconstructed or non-prompt leptons,
therefore different data-based corrections are applied.
The \ttW\ background is the most important background in the \SSML\ channel, and
a detailed  model of systematic uncertainties
was used, while in the \SLOS\ channel it is a small background.
Uncertainties in \ttW\ and \ttjets\ backgrounds are thus uncorrelated.

The following sections present the results from the \SLOS\ final states only and those from the combination of the \SLOS\ and \SSML\ final states.

\subsection{Limits on \ttHAtt\ production in a 2HDM}
\label{sub:limits}

The expected and observed 95\% CL upper limits on the \ttHAtt\ production cross-sections
as a function of $m_{H/A}$ mass are shown in Figure~\ref{fig:data_SplusB_limits}.
The theoretical predictions at NLO in QCD for two different values of $\tan\beta$ are also
shown~\cite{Harlander:2012pb,Harlander:2002wh,Harlander:2003ai,Aglietti:2004nj,Bonciani:2010ms,Harlander:2005rq}.
Overall, the observed limits are consistent with the expected ones. %
Similar to the results in the \SSML\ final states~\cite{EXOT-2019-26},
there is a small excess in the observed limits compared with the expected ones.
The mass point with the largest deviation is $m_{H/A} = 500$~\GeV.
The local significance of the fitted signal at this point is evaluated to be 2.1 standard deviations.
All other mass points have a local significance of less than 1.5 standard deviations.
The goodness-of-fit under the background-only hypothesis is evaluated for all the mass points,
using a likelihood-ratio test in which the nominal fit is compared with that of a saturated model~\cite{SaturatedModel}.
The smallest goodness-of-fit value of 49\% is observed for $m_{H/A} = 800$~\GeV,
whilst for $m_{H/A} = 500$~\GeV\ the value is 64\%.
Other mass points have a goodness-of-fit value of greater than 81\%.

\begin{figure}[t!]
\centering
\includegraphics[width=0.6\textwidth]{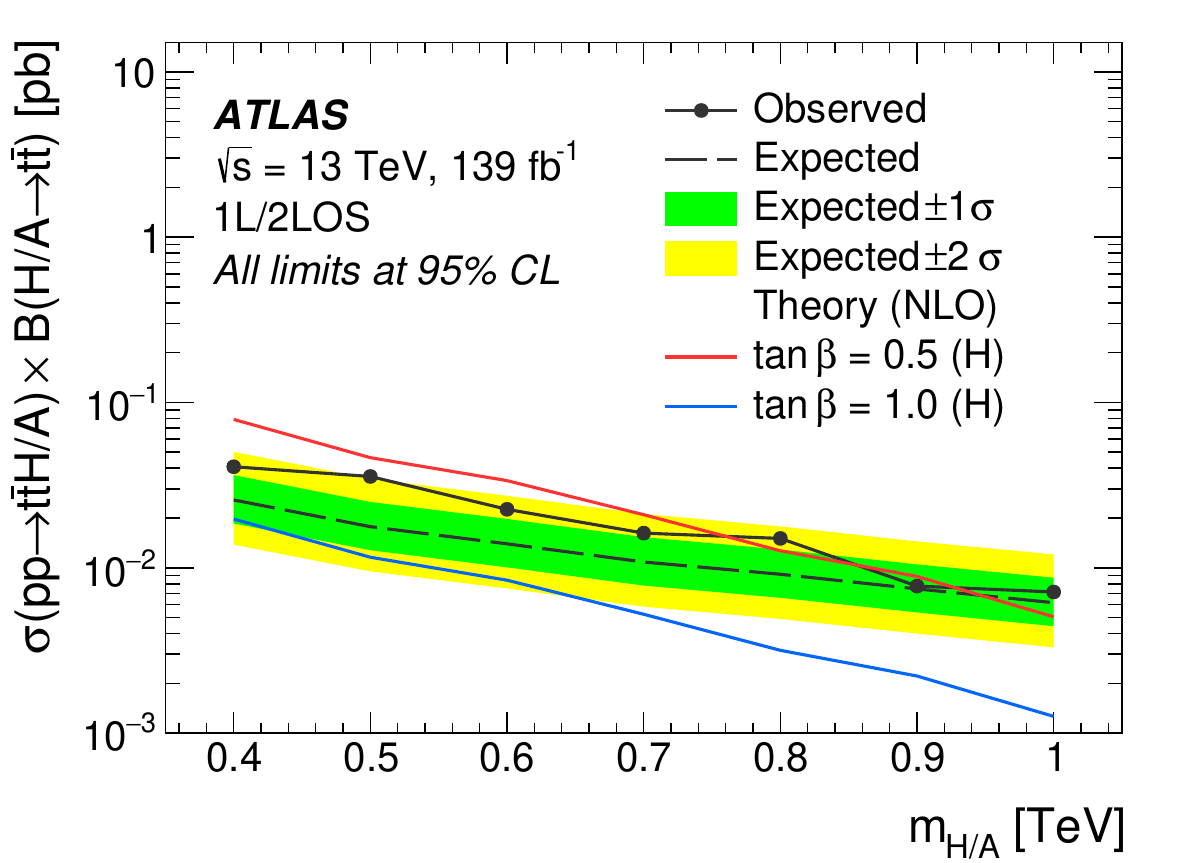}
\caption{Observed (solid line) and expected (dashed line) 95\% CL upper limits
on the cross-section times branching ratio,
$\sigma(pp\rightarrow t\bar{t}H/A) \times B(H/A \rightarrow t\bar{t})$,
as a function of $m_{H/A}$. The green (yellow) band illustrates the $\pm1\sigma$ ($\pm2\sigma$)
interval around the expected limit.
Theoretical predictions for two values of $\tan\beta$ are shown for an $H$ boson.
The predicted values for $A$ differ by less than 3\textperthousand\ for all mass points for the two $\tan\beta$ values shown.
}
\label{fig:data_SplusB_limits}
\end{figure}

The upper limits on the cross-sections are interpreted as exclusion limits on $\tan\beta$
in the context of a 2HDM Type-II model in the alignment limit.
As described in Section~\ref{sec:mc},
the signal kinematics is the same at different $\tan\beta$ values and for $H$ and $A$.
Therefore, the limits on $\tan\beta$ are obtained by
comparing the cross-section upper limits in Figure~\ref{fig:data_SplusB_limits}
with the predicted signal cross-section for \ttHtt\ and \ttAtt\ at different values of $\tan\beta$.
The uncertainty in the theory cross-sections arises from
variations of the PDFs and the renormalisation and factorisation scales.
The expected and observed 95\% CL lower limits for $\tan\beta$ as a function of the
$m_{H/A}$ mass for different scenarios are shown in Figure~\ref{fig:data_tanB_limits}.
Three scenarios are shown where
(a) both the scalar $H$ and pseudo-scalar $A$ bosons contribute to the $t\bar{t}t\bar{t}$ cross-section with equal masses,
(b) only an $H$ boson contributes,
and (c) only an $A$ boson contributes.

\begin{figure}[t!]
\centering
\subfloat[]{\includegraphics[width=0.495\textwidth]{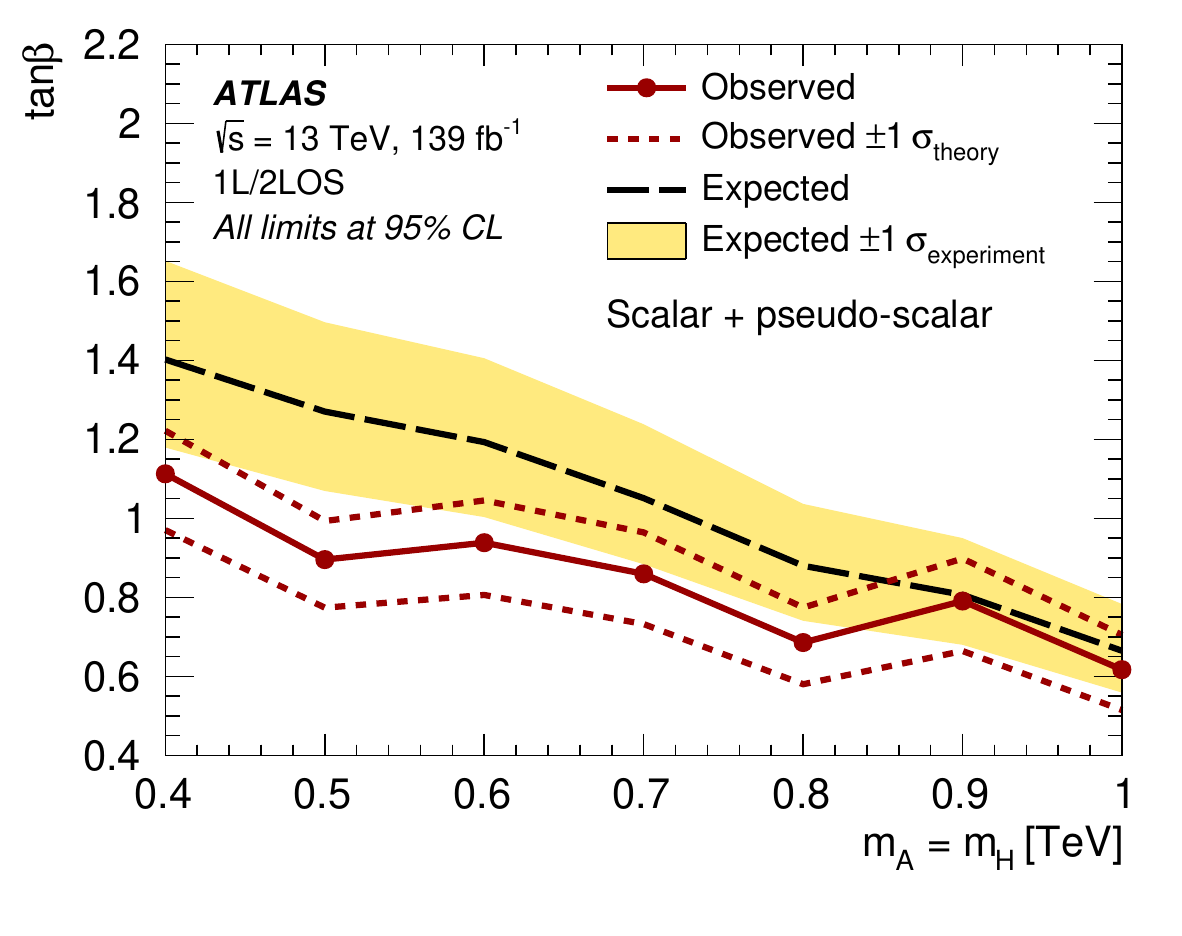}}\\
\subfloat[]{\includegraphics[width=0.495\textwidth]{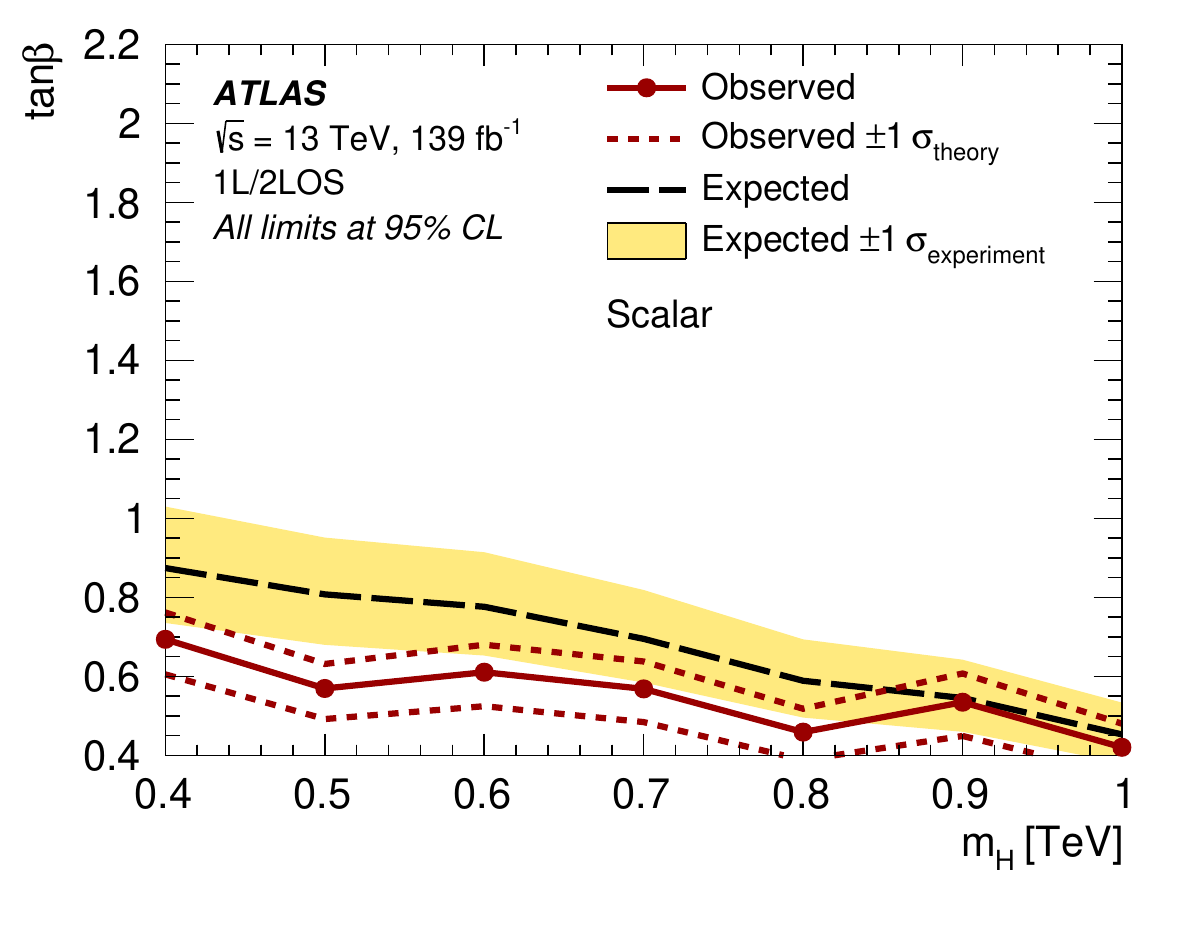}}
\subfloat[]{\includegraphics[width=0.495\textwidth]{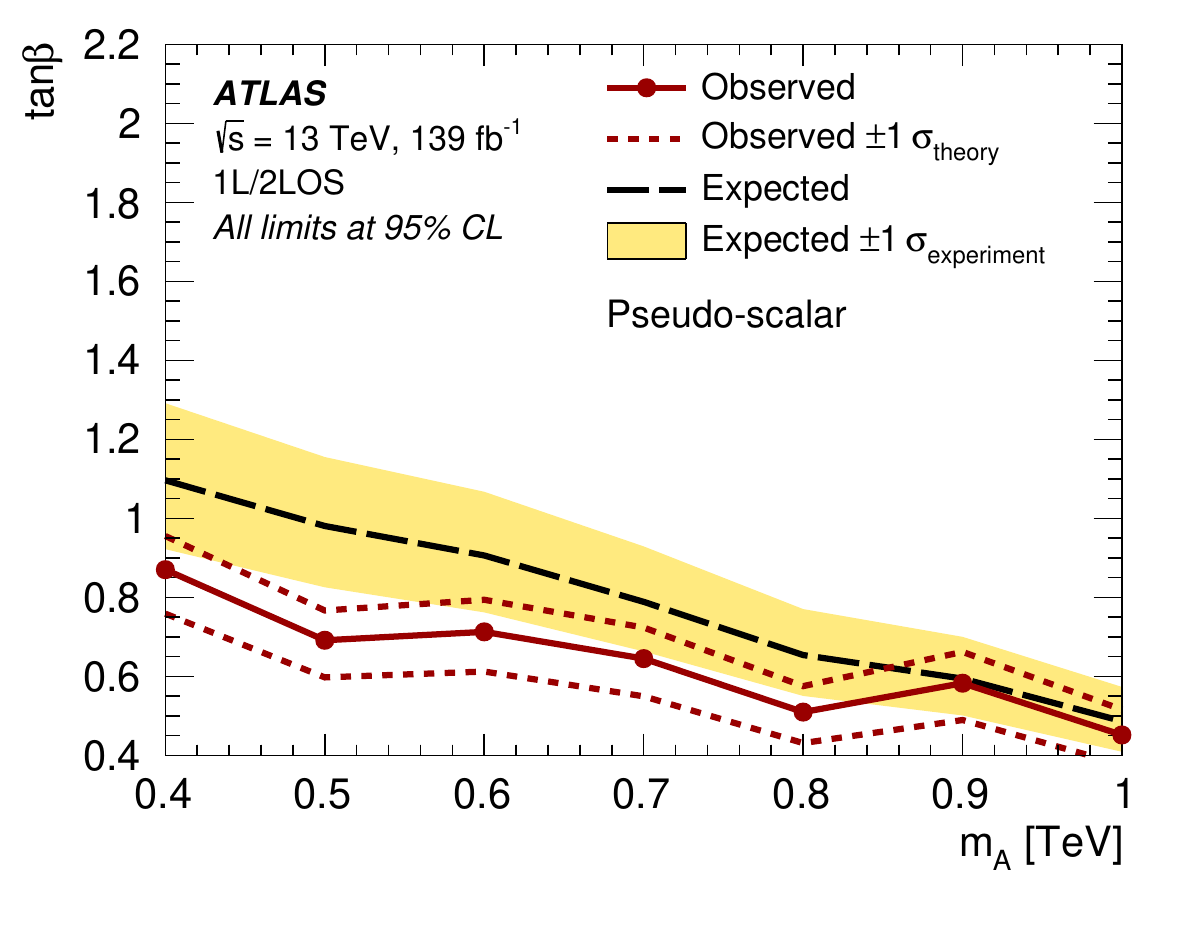}}
\caption{Observed (red solid line) and expected (black dashed line) 95\% CL lower limits
on $\tan\beta$ as a function of the $m_{H/A}$ mass
obtained using the \SLOS\ final states,
assuming a Type-II 2HDM in the alignment limit.
Values of $\tan\beta$ below the observed limit are excluded.
The scenario where both the scalar $H$ and pseudo-scalar $A$ contribute with $m_H = m_A$ is shown in (a).
The scenarios with the contribution from $H$ and $A$ only are shown in (b) and (c), respectively.
The yellow bands illustrate the $\pm1\sigma$ intervals around the expected limits.
The two red dashed lines correspond to the observed limits
when the theory cross-sections are varied by $\pm1\sigma$.
Values of  $\tan\beta$ $< 0.4$ are not considered so that the top-quark Yukawa coupling
remains in the perturbative regime~\cite{Branco:2011iw}.}
\label{fig:data_tanB_limits}
\end{figure}

The impacts of the systematic and statistical uncertainties in the fitted \ttHAtt\ cross-sections
are given in Table~\ref{tab:grouped_impact} for three illustrative $m_{H/A}$ values.
The impacts are evaluated using the post-fit uncertainties in the \ttHAtt\ cross-sections
from the fit to data under the signal-plus-background hypothesis,
and can be compared with the cross-section upper limits shown in Figure~\ref{fig:data_SplusB_limits}.
The impacts are grouped into categories according to their origin.
As mentioned in Section~\ref{sec:systematics} the largest sources of systematic errors arise from
the group of uncertainties in the modelling of the \ttjets\ background.
The leading individual uncertainties include those from: the comparison between the 4FS and 5FS predictions
and the PS, hadronisation and fragmentation modelling in \ttbb\ and \ttbbb\ background components;
the generator comparison of SM \tttt\ production;
the normalisation of \ttcjets\ and \ttbjets;
and the NN kinematic corrections in the \OS\ channel.

\begin{table}[t!]
\begin{center}
\caption{The list of impacts
on the fitted \ttHAtt\ cross-sections
from all sources of uncertainties,
for $m_{H/A} = 400$~\GeV, 700~\GeV\ and 1000~\GeV,
grouped in categories according to their origin.
The impacts are evaluated using the post-fit uncertainties in the \ttHAtt\ cross-sections (the total uncertainty)
from the fit to data under the signal-plus-background hypothesis including all uncertainties (referred to as the nominal fit).
The statistical uncertainty is evaluated as the post-fit uncertainty in the \ttHAtt\ cross-sections
when fixing all systematic uncertainties to their post-fit values from the nominal fit.
The impacts of systematic uncertainties are shown for groups of individual uncertainties of similar nature.
The value for each group is obtained by fixing the group of uncertainties to their post-fit values from the nominal fit,
and taking the quadratic difference between the resulting uncertainty in the \ttHAtt\ cross-section
and the total uncertainty.
The total systematic uncertainty is the quadratic difference between the total uncertainty and the statistical uncertainty.
The sum in quadrature of all systematic uncertainties can be different from the total systematic uncertainty in the table due to
the correlations amongst the uncertainties.
}
\label{tab:grouped_impact}
\begin{tabular}{l|cc|cc|cc}
\toprule \midrule
Uncertainty source & \multicolumn{6}{c}{$\Delta\sigma_{\ttHAtt}$ [fb]} \\
& \multicolumn{2}{c}{$m_{H/A}$=400~\GeV} & \multicolumn{2}{c}{$m_{H/A}$=700~\GeV} & \multicolumn{2}{c}{$m_{H/A}$=1000~\GeV} \\
\midrule
\textbf{Signal Modelling} & & & & & & \\
BSM $t\bar{t}t\bar{t}$ modelling & \multicolumn{2}{c|}{$ <1$} & +0.1 & $ < 0.1$ & \multicolumn{2}{c}{$ < 0.1$} \\
\midrule
\textbf{Background Modelling} & & & & & & \\
\ttbjets\ modelling & +11 & -10 & +3.7 & -3.4 & +1.9 & -1.7 \\
SM \tttt\ modelling & +3 & -3 & +2.1 & -2.1 & +0.9 & -0.9 \\
\ttjets\ reweighting & +3 & -3 & +1.0 & -1.0 & +0.5 & -0.5 \\
\ttcjets\ modelling & +2 & -2 & +0.9 & -0.8 & +0.4 & -0.4 \\
\ttljets\ modelling & +1 & -1 & +0.2 & -0.2 & \multicolumn{2}{c}{$ < 0.1$} \\
Other background modelling & \multicolumn{2}{c|}{$ < 1$} & +0.4 & -0.4 & +0.2 & -0.2 \\
\midrule
\textbf{Experimental} & & & & & & \\
Jet energy scale and resolution & +4 & -2 & +1.3 & -0.8 & +0.5 & -0.3 \\
MC statistical uncertainties & +2 & -3 & +0.6 & -0.7 & +0.4 & -0.4 \\
$b$-tagging efficiency & +2 & -1 & +0.7 & -0.4 & +0.4 & -0.4 \\
Other uncertainties & \multicolumn{2}{c|}{$< 1$} & +0.3 & -0.5 & +0.1 & -0.2 \\
Luminosity & \multicolumn{2}{c|}{$ < 1$} & +0.3 & -0.1 & \multicolumn{2}{c}{$ < 0.1$} \\
\midrule \midrule
\textbf{Total systematic uncertainty} & +13 & -12 & +4.8 & -4.6 & +2.5 & -2.4 \\
\midrule \midrule
\textbf{Statistical uncertainty} & +6 & -6 & +3.3 & -3.2 & +2.3 & -2.2 \\
\midrule \midrule
\textbf{Total uncertainty} & +14 & -13 & +5.6 & -5.4 & +3.2 & -3.0 \\
\midrule \bottomrule
\end{tabular}
\end{center}
\end{table}

Figure~\ref{fig:xsec_combination_limit} shows the observed and expected 95\% CL upper limits on
the \ttHAtt\ production cross-section from the combined fit to all SRs and CRs in the \SLOS\ and \SSML\ analyses.
The observed (expected) upper limits range from 14 (9.4)~fb for $m_{H/A} = 400$~\GeV\ to 5.0 (3.8)~fb for $m_{H/A} = 1000$~\GeV.
The figure also compares the expected upper limits from the \SLOS\ and \SSML\ with the one from the combination.
The sensitivity in the low mass regime is dominated by the \SSML\ final states.
The combination improves the sensitivity relative to using only the \SSML\ final states
in the high mass regime.
The largest improvement in the observed upper limit is 19\% for $m_{H/A} = 900$~\GeV,
with the expected upper limits across the full mass range improving by up to 18\%.

The resulting 95\% CL lower limits for $\tan\beta$ from the combination are shown in Figure~\ref{fig:tanB_combination_limit}.
For the case where $m_A = m_H$, at low masses, this result is less restrictive
than that from the $gg\rightarrow A/H \rightarrow \ttbar$ search~\cite{EXOT-2020-25}.
At high masses, the expected sensitivity is comparable.

\begin{figure}[t!]
\centering
\includegraphics[width=0.6\textwidth]{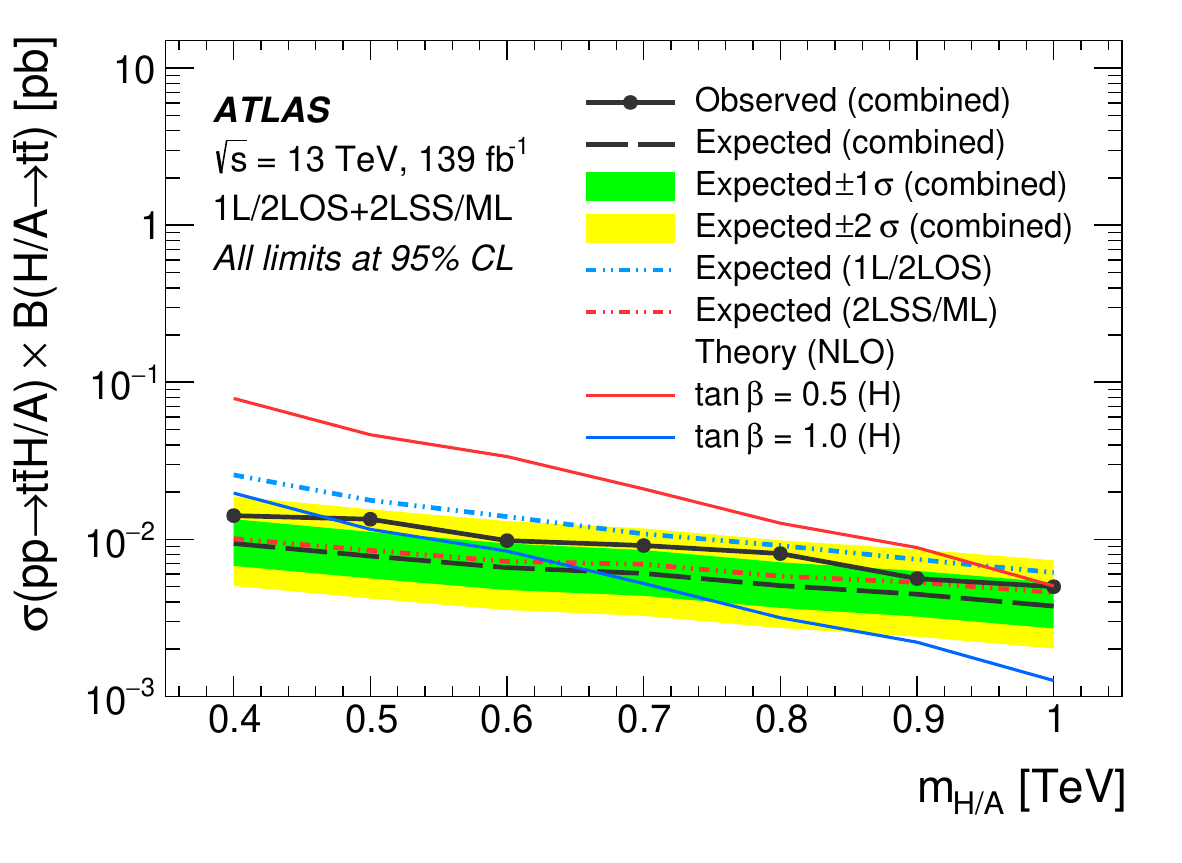}
\caption{Expected and observed 95\% CL upper limits
on the cross-section times branching ratio,
$\sigma(pp\rightarrow t\bar{t}H/A) \times B(H/A \rightarrow t\bar{t})$,
as a function of $m_{H/A}$,
obtained from the combination of the \SLOS\ and \SSML\ final states.
The green (yellow) band illustrates the $\pm1\sigma$ ($\pm2\sigma$)
interval around the expected limit from the combination.
The expected limits from the individual \SLOS\ and \SSML\ analyses are also shown.
Theoretical predictions for two values of $\tan\beta$ are shown for an $H$ boson.
The predicted values for $A$ differ by less than 3\textperthousand\ for all mass points for the two $\tan\beta$ values shown.
}
\label{fig:xsec_combination_limit}
\end{figure}

\begin{figure}[t!]
\centering
\subfloat[]{\includegraphics[width=0.495\textwidth]{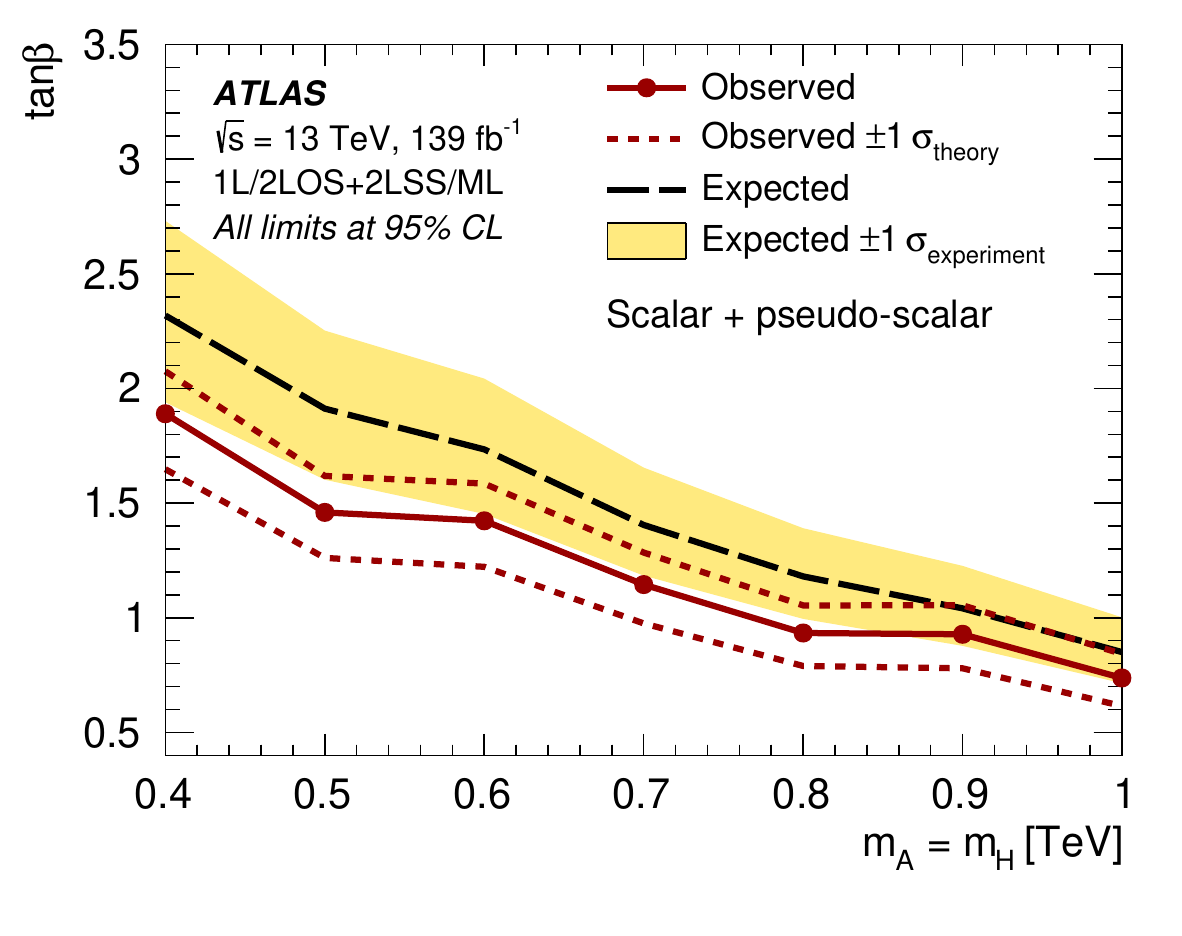}}\\
\subfloat[]{\includegraphics[width=0.495\textwidth]{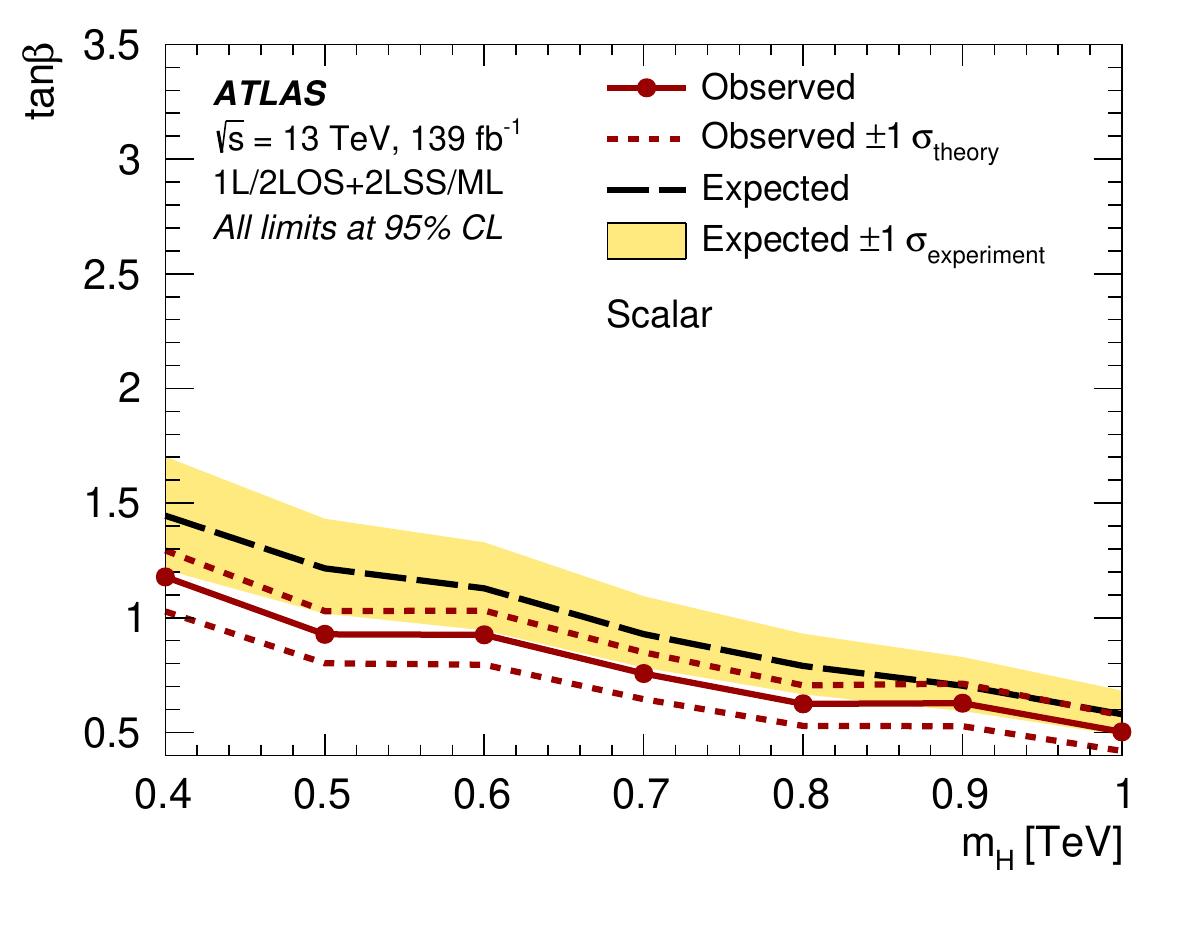}}
\subfloat[]{\includegraphics[width=0.495\textwidth]{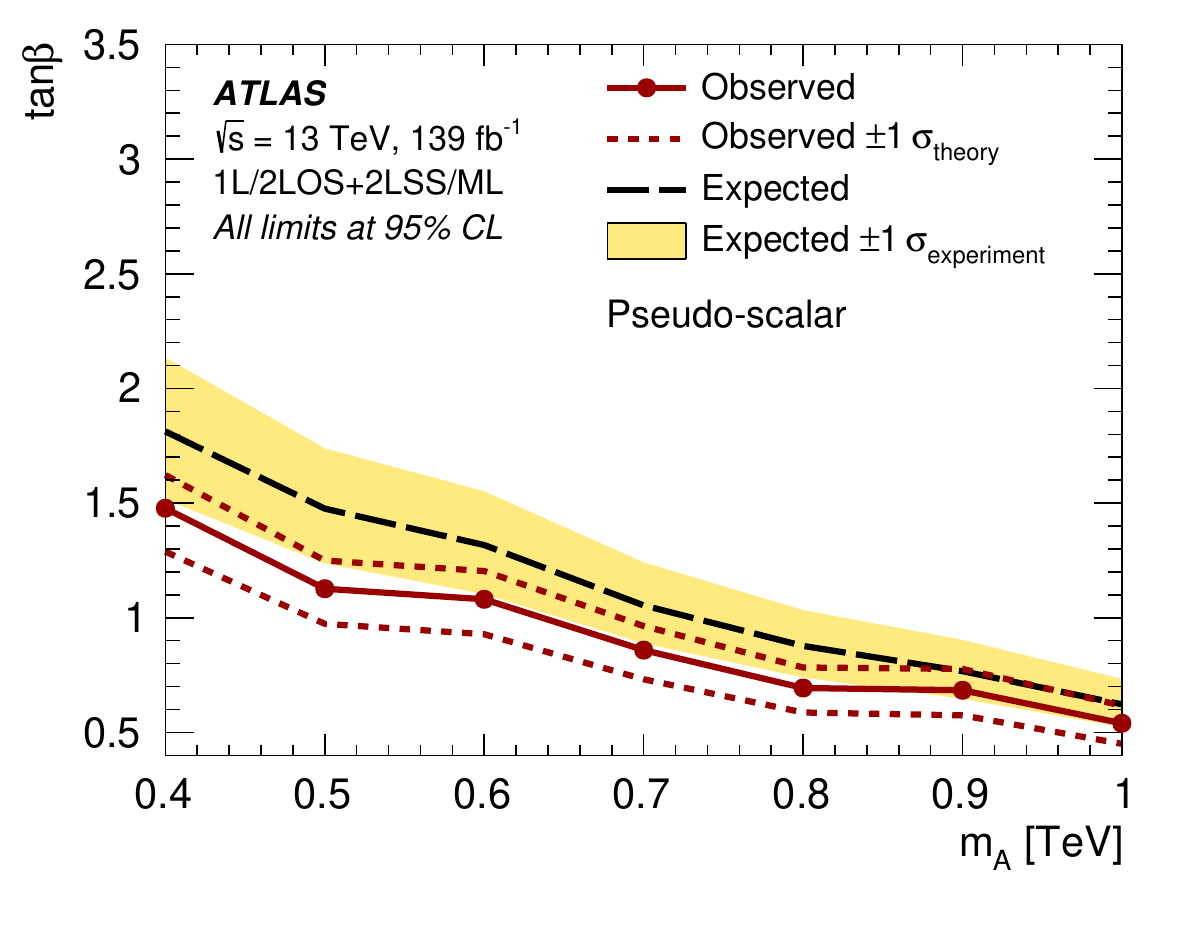}}
\caption{Expected (black dashed line) and observed (red solid line) 95\% CL limits
on $\tan\beta$ as a function of the $m_{H/A}$ mass
obtained using the combination of \SLOS\ and \SSML\ final states,
assuming type-II 2HDM in the alignment limit.
Values of $\tan\beta$ below the observed limit are excluded.
The scenario where both the scalar $H$ and pseudo-scalar $A$ contribute with $m_H = m_A$ is shown in (a).
The scenarios with the contribution from $H$ and $A$ only are shown in (b) and (c), respectively.
The yellow bands illustrate the $\pm1\sigma$ bands of the expected limits.
The two red dashed lines correspond to the observed limits
when the theory cross-sections are varied by $\pm1\sigma$.
Values of  $\tan\beta$ $< 0.4$
are not considered so that the top-quark Yukawa coupling
remains in the perturbative regime~\cite{Branco:2011iw}.}
\label{fig:tanB_combination_limit}
\end{figure}

\subsection{Limits on \texorpdfstring{$S_8S_8\rightarrow\tttt$}{S8S8->tttt} production}
\label{sub:sgluon_limits}

The same analysis strategy is used to derive upper limits on the $S_8S_8\rightarrow\tttt$ production cross-section.
The \SLOS\ and \SSML\ final states are combined to determine the final limits.
For the mass points with $m_{S_8} \leq 1$~\TeV, the $m_{H/A}$-parameterised classifiers (GNN for \SLOS\ and BDT for \SSML)
are evaluated at $m_{S_8} = m_{H/A}$.
At each mass point, the binning of the fitted distributions is the same as the one used for $H/A$.
For $m_{S_8} > 1$~\TeV, the GNN/BDT and binning for $m_{H/A} = 1$~\TeV\ are adopted.
The resulting 95\% CL upper limits on the $S_8S_8\rightarrow\tttt$ production cross-section are illustrated in Figure~\ref{fig:sgluon_limits}.
The observed and expected upper limits from the combination of \SLOS\ and \SSML\ final states are shown,
compared with the expected limits from using only the \SLOS\ or the \SSML\ final states.
The behaviour of the limits for $m_{S_8} \leq 1$~\TeV\ is similar to that for \ttHAtt\ production,
with an improvement of up to 26\% when using the combination relative to using the \SSML\ final states only.
The limits slightly weaken for $m_{S_8} > 1$~\TeV\ given that the classifiers are optimised for $m_{H/A} = 1$~\TeV.
For the sgluon signals under consideration as described in Section~\ref{sec:mc},
$m_{S_8} \leq 1.3$~\TeV\ are excluded at 95\% CL.

\begin{figure}[t!]
\centering
\includegraphics[width=0.6\textwidth]{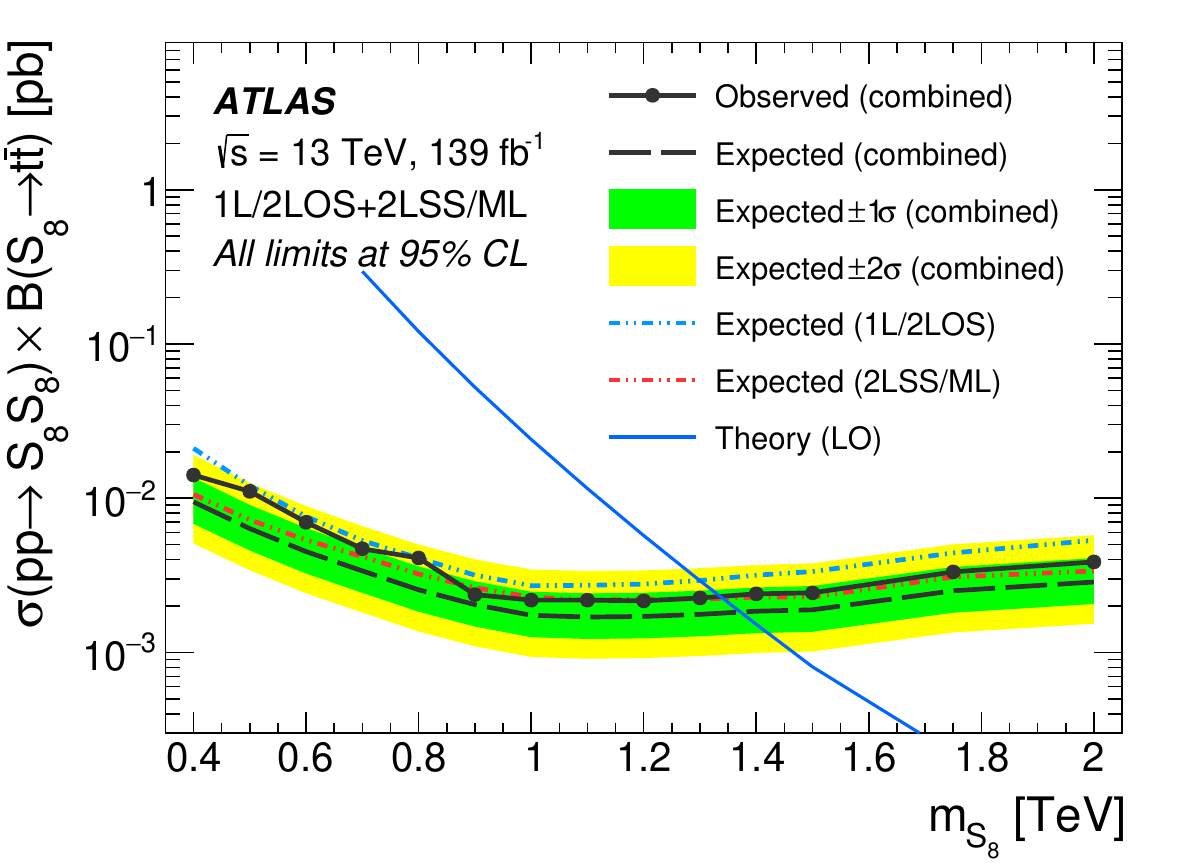}
\caption{Expected and observed 95\% CL upper limits
on the $S_8S_8\rightarrow\tttt$ production cross-section
as a function of $m_{S_8}$,
obtained from the combination of the \SLOS\ and \SSML\ final states.
The expected limits from the individual \SLOS\ and \SSML\ analyses are also shown.
The predicted production cross-section from Ref.~\cite{Darm__2021} is shown with the solid blue curve.
}
\label{fig:sgluon_limits}
\end{figure}


%
%
\FloatBarrier

%
\section{Conclusions}
\label{sec:conclusion}

A search for \ttHAtt\ production
is presented using the full LHC Run~2 data sample collected by the ATLAS experiment
during 2015--2018, corresponding to an integrated luminosity of 139~\ifb of
$pp$ collisions at $\sqrt{s} = 13$~\TeV.
Events with one lepton or two leptons with opposite-sign charges and high jet multiplicities are selected.
The dominant background in all selected regions is \ttjets.
Data-driven corrections are applied to the normalisation of the \ttljets, \ttcjets\ and \ttbjets\ background components,
and their kinematics.
The kinematic corrections are derived based on NNs
and improve the modelling of the \ttjets\ background in multiple dimensions.
A $m_{H/A}$-parameterised GNN is used to optimise the signal-to-background discrimination.

Upper limits at 95\% CL are set on the production cross-section times branching ratio,
$\sigma(pp\rightarrow t\bar{t}H/A) \times B(H/A \rightarrow t\bar{t})$,
for $H/A$ masses between 400~\GeV\ and 1000~\GeV.  These results are combined
with a previous search from ATLAS using multilepton final states.
The combined upper limits range from
14~fb at 400~\GeV\ to 5.0~fb at 1000~\GeV,
with an improvement of up to 19\% relative to using only the multilepton final states.
These results are interpreted in the context of a 2HDM Type-II model in the alignment limit
to set limits on the values of $\tan\beta$.
Assuming only the scalar $H$ (pseudo-scalar $A$) boson
contributes to the cross-section, values of $\tan\beta$ below 1.2 (1.5) and 0.5 (0.5) are
excluded at the low and high mass $H/A$ mass values.
In case both the particles contribute, the exclusion range
is below 1.9 and 0.7 at masses of 400~\GeV\ and 1000~\GeV.
When considering the sgluon pair production signals with $m_{S_8}$ from 400~\GeV\ to 2~\TeV,
the observed 95\% CL upper limits on the $S_8S_8 \rightarrow \tttt$ cross-section
range from 14~fb (400~\GeV) to 2.2~fb (1~\TeV).
Sgluon masses $m_{S_8} \leq 1.3$~\TeV\ are excluded.


%
%
\section*{Acknowledgements}
%

%
%

%
%

We thank CERN for the very successful operation of the LHC and its injectors, as well as the support staff at
CERN and at our institutions worldwide without whom ATLAS could not be operated efficiently.

The crucial computing support from all WLCG partners is acknowledged gratefully, in particular from CERN, the ATLAS Tier-1 facilities at TRIUMF/SFU (Canada), NDGF (Denmark, Norway, Sweden), CC-IN2P3 (France), KIT/GridKA (Germany), INFN-CNAF (Italy), NL-T1 (Netherlands), PIC (Spain), RAL (UK) and BNL (USA), the Tier-2 facilities worldwide and large non-WLCG resource providers. Major contributors of computing resources are listed in Ref.~\cite{ATL-SOFT-PUB-2023-001}.

We gratefully acknowledge the support of ANPCyT, Argentina; YerPhI, Armenia; ARC, Australia; BMWFW and FWF, Austria; ANAS, Azerbaijan; CNPq and FAPESP, Brazil; NSERC, NRC and CFI, Canada; CERN; ANID, Chile; CAS, MOST and NSFC, China; Minciencias, Colombia; MEYS CR, Czech Republic; DNRF and DNSRC, Denmark; IN2P3-CNRS and CEA-DRF/IRFU, France; SRNSFG, Georgia; BMBF, HGF and MPG, Germany; GSRI, Greece; RGC and Hong Kong SAR, China; ISF and Benoziyo Center, Israel; INFN, Italy; MEXT and JSPS, Japan; CNRST, Morocco; NWO, Netherlands; RCN, Norway; MNiSW, Poland; FCT, Portugal; MNE/IFA, Romania; MESTD, Serbia; MSSR, Slovakia; ARIS and MVZI, Slovenia; DSI/NRF, South Africa; MICIU/AEI, Spain; SRC and Wallenberg Foundation, Sweden; SERI, SNSF and Cantons of Bern and Geneva, Switzerland; NSTC, Taipei; TENMAK, T\"urkiye; STFC/UKRI, United Kingdom; DOE and NSF, United States of America.

Individual groups and members have received support from BCKDF, CANARIE, CRC and DRAC, Canada; PRIMUS 21/SCI/017, CERN-CZ and FORTE, Czech Republic; COST, ERC, ERDF, Horizon 2020, ICSC-NextGenerationEU and Marie Sk{\l}odowska-Curie Actions, European Union; Investissements d'Avenir Labex, Investissements d'Avenir Idex and ANR, France; DFG and AvH Foundation, Germany; Herakleitos, Thales and Aristeia programmes co-financed by EU-ESF and the Greek NSRF, Greece; BSF-NSF and MINERVA, Israel; Norwegian Financial Mechanism 2014-2021, Norway; NCN and NAWA, Poland; La Caixa Banking Foundation, CERCA Programme Generalitat de Catalunya and PROMETEO and GenT Programmes Generalitat Valenciana, Spain; G\"{o}ran Gustafssons Stiftelse, Sweden; The Royal Society and Leverhulme Trust, United Kingdom.

In addition, individual members wish to acknowledge support from CERN: European Organization for Nuclear Research (CERN PJAS); Chile: Agencia Nacional de Investigaci\'on y Desarrollo (FONDECYT 1190886, FONDECYT 1210400, FONDECYT 1230812, FONDECYT 1230987); China: Chinese Ministry of Science and Technology (MOST-2023YFA1605700, MOST-2023YFA1609300), National Natural Science Foundation of China (NSFC - 12175119, NSFC 12275265, NSFC-12075060); Czech Republic: Czech Science Foundation (GACR - 24-11373S), Ministry of Education Youth and Sports (FORTE CZ.02.01.01/00/22\_008/0004632), PRIMUS Research Programme (PRIMUS/21/SCI/017); EU: H2020 European Research Council (ERC - 101002463); European Union: European Research Council (ERC - 948254, ERC 101089007), Horizon 2020 Framework Programme (MUCCA - CHIST-ERA-19-XAI-00), European Union, Future Artificial Intelligence Research (FAIR-NextGenerationEU PE00000013), Italian Center for High Performance Computing, Big Data and Quantum Computing (ICSC, NextGenerationEU); France: Agence Nationale de la Recherche (ANR-20-CE31-0013, ANR-21-CE31-0013, ANR-21-CE31-0022, ANR-22-EDIR-0002), Investissements d'Avenir Labex (ANR-11-LABX-0012); Germany: Baden-Württemberg Stiftung (BW Stiftung-Postdoc Eliteprogramme), Deutsche Forschungsgemeinschaft (DFG - 469666862, DFG - CR 312/5-2); Italy: Istituto Nazionale di Fisica Nucleare (ICSC, NextGenerationEU), Ministero dell'Università e della Ricerca (PRIN - 20223N7F8K - PNRR M4.C2.1.1); Japan: Japan Society for the Promotion of Science (JSPS KAKENHI JP21H05085, JSPS KAKENHI JP22H01227, JSPS KAKENHI JP22H04944, JSPS KAKENHI JP22KK0227); Netherlands: Netherlands Organisation for Scientific Research (NWO Veni 2020 - VI.Veni.202.179); Norway: Research Council of Norway (RCN-314472); Poland: Ministry of Science and Higher Education (IDUB AGH, POB8, D4 no 9722), Polish National Agency for Academic Exchange (PPN/PPO/2020/1/00002/U/00001), Polish National Science Centre (NCN 2021/42/E/ST2/00350, NCN OPUS nr 2022/47/B/ST2/03059, NCN UMO-2019/34/E/ST2/00393, NCN \& H2020 MSCA 945339, UMO-2020/37/B/ST2/01043, UMO-2021/40/C/ST2/00187, UMO-2022/47/O/ST2/00148, UMO-2023/49/B/ST2/04085); Slovenia: Slovenian Research Agency (ARIS grant J1-3010); Spain: Generalitat Valenciana (Artemisa, FEDER, IDIFEDER/2018/048), Ministry of Science and Innovation (MCIN \& NextGenEU PCI2022-135018-2, MICIN \& FEDER PID2021-125273NB, RYC2019-028510-I, RYC2020-030254-I, RYC2021-031273-I, RYC2022-038164-I), PROMETEO and GenT Programmes Generalitat Valenciana (CIDEGENT/2019/023, CIDEGENT/2019/027); Sweden: Carl Trygger Foundation (Carl Trygger Foundation CTS 22:2312), Swedish Research Council (Swedish Research Council 2023-04654, VR 2018-00482, VR 2022-03845, VR 2022-04683, VR 2023-03403, VR grant 2021-03651), Knut and Alice Wallenberg Foundation (KAW 2018.0157, KAW 2018.0458, KAW 2019.0447, KAW 2022.0358); Switzerland: Swiss National Science Foundation (SNSF - PCEFP2\_194658); United Kingdom: Leverhulme Trust (Leverhulme Trust RPG-2020-004), Royal Society (NIF-R1-231091); United States of America: U.S. Department of Energy (ECA DE-AC02-76SF00515), Neubauer Family Foundation.

%
%


%
\clearpage
\printbibliography
\clearpage
 
\begin{flushleft}
\hypersetup{urlcolor=black}
{\Large The ATLAS Collaboration}

\bigskip

\AtlasOrcid[0000-0002-6665-4934]{G.~Aad}$^\textrm{\scriptsize 104}$,
\AtlasOrcid[0000-0001-7616-1554]{E.~Aakvaag}$^\textrm{\scriptsize 17}$,
\AtlasOrcid[0000-0002-5888-2734]{B.~Abbott}$^\textrm{\scriptsize 123}$,
\AtlasOrcid[0000-0002-0287-5869]{S.~Abdelhameed}$^\textrm{\scriptsize 119a}$,
\AtlasOrcid[0000-0002-1002-1652]{K.~Abeling}$^\textrm{\scriptsize 56}$,
\AtlasOrcid[0000-0001-5763-2760]{N.J.~Abicht}$^\textrm{\scriptsize 50}$,
\AtlasOrcid[0000-0002-8496-9294]{S.H.~Abidi}$^\textrm{\scriptsize 30}$,
\AtlasOrcid[0009-0003-6578-220X]{M.~Aboelela}$^\textrm{\scriptsize 45}$,
\AtlasOrcid[0000-0002-9987-2292]{A.~Aboulhorma}$^\textrm{\scriptsize 36e}$,
\AtlasOrcid[0000-0001-5329-6640]{H.~Abramowicz}$^\textrm{\scriptsize 155}$
% ...
\bigskip
\\

% ...
$^{17}$Department for Physics and Technology, University of Bergen, Bergen; Norway.\\
% ...
$^{30}$Physics Department, Brookhaven National Laboratory, Upton NY; United States of America.\\
% ...
$^{36}$$^{(a)}$Facult\'e des Sciences Ain Chock, Universit\'e Hassan II de Casablanca;$^{(b)}$Facult\'{e} des Sciences, Universit\'{e} Ibn-Tofail, K\'{e}nitra;$^{(c)}$Facult\'e des Sciences Semlalia, Universit\'e Cadi Ayyad, LPHEA-Marrakech;$^{(d)}$LPMR, Facult\'e des Sciences, Universit\'e Mohamed Premier, Oujda;$^{(e)}$Facult\'e des sciences, Universit\'e Mohammed V, Rabat;$^{(f)}$Institute of Applied Physics, Mohammed VI Polytechnic University, Ben Guerir; Morocco.\\
% ...
$^{45}$Physics Department, Southern Methodist University, Dallas TX; United States of America.\\
% ...
$^{50}$Fakult\"{a}t Physik , Technische Universit{\"a}t Dortmund, Dortmund; Germany.\\
% ...
$^{56}$II. Physikalisches Institut, Georg-August-Universit\"{a}t G\"ottingen, G\"ottingen; Germany.\\
% ...
$^{104}$CPPM, Aix-Marseille Universit\'e, CNRS/IN2P3, Marseille; France.\\
% ...
$^{119}$$^{(a)}$New York University Abu Dhabi, Abu Dhabi;$^{(b)}$United Arab Emirates University, Al Ain; United Arab Emirates.\\
% ...
$^{123}$Homer L. Dodge Department of Physics and Astronomy, University of Oklahoma, Norman OK; United States of America.\\
% ...
$^{155}$Raymond and Beverly Sackler School of Physics and Astronomy, Tel Aviv University, Tel Aviv; Israel.\\
% ...
$^{a}$ Also Affiliated with an institute covered by a cooperation agreement with CERN.\\
$^{b}$ Also at An-Najah National University, Nablus; Palestine.\\
$^{c}$ Also at Borough of Manhattan Community College, City University of New York, New York NY; United States of America.\\
$^{d}$ Also at Center for High Energy Physics, Peking University; China.\\
$^{e}$ Also at Center for Interdisciplinary Research and Innovation (CIRI-AUTH), Thessaloniki; Greece.\\
$^{f}$ Also at Centro Studi e Ricerche Enrico Fermi; Italy.\\
% ...

\end{flushleft}


%

%
%

%

%
%
%
%

\end{document}